\documentclass[prd,eqsecnum,twocolumn,amsfonts,amssymb]{revtex4}

\usepackage{CJK}

\usepackage{graphicx}

\usepackage{bm}

\newcommand{\beq}{\begin{equation}}
\newcommand{\eeq}{\end{equation}}
\newcommand{\beqs}{\begin{eqnarray}}
\newcommand{\eeqs}{\end{eqnarray}}

\newcommand{\Dslash}{D\hspace{-0.09in}\slash}
\newcommand{\drawsquare}[2]{\hbox{%
\rule{#2pt}{#1pt}\hskip-#2pt%  left vertical
\rule{#1pt}{#2pt}\hskip-#1pt%  lower horizontal
\rule[#1pt]{#1pt}{#2pt}}\rule[#1pt]{#2pt}{#2pt}\hskip-#2pt%  upper horizontal
\rule{#2pt}{#1pt}}% right vertical
\newcommand{\fund}{\raisebox{-.5pt}{\drawsquare{6.5}{0.4}}}%  fund
\newcommand{\sym}{\raisebox{-.5pt}{\drawsquare{6.5}{0.4}}\hskip-0.4pt%
        \raisebox{-.5pt}{\drawsquare{6.5}{0.4}}}%  symmetric second rank
\newcommand{\asym}{\raisebox{-3.5pt}{\drawsquare{6.5}{0.4}}\hskip-6.9pt%
        \raisebox{3pt}{\drawsquare{6.5}{0.4}}}%  antisymmetric second rank

\begin{document}

\begin{CJK*}{UTF8}{}

\title{Dynamical Symmetry Breaking in Chiral Gauge Theories with 
Direct-Product Gauge Groups} 

\author{Yan-Liang Shi (\CJKfamily{bsmi}石炎亮) and Robert Shrock}

\affiliation{C. N. Yang Institute for Theoretical Physics, 
Stony Brook University, Stony Brook, N. Y. 11794 }

\begin{abstract}

We analyze patterns of dynamical symmetry breaking in strongly coupled
chiral gauge theories with direct-product gauge groups $G$. 
If the gauge coupling for a factor group $G_i \subset G$ becomes 
sufficiently strong, it can produce bilinear fermion condensates that break 
the $G_i$ symmetry itself and/or break other gauge symmetries $G_j \subset G$.
Our comparative study of a number of strongly coupled direct-product chiral
gauge theories elucidates how the patterns of symmetry breaking depend on the
structure of $G$ and on the relative sizes of the gauge 
couplings corresponding to factor groups in the direct product. 

\end{abstract}

\maketitle

\end{CJK*}

% =======================================================================

\section{Introduction}
\label{intro_section}

A problem of longstanding interest has been the behavior of strongly coupled
chiral gauge theories (in four spacetime dimensions, at zero temperature).
Here a chiral gauge theory is defined as one in which the fermions, written in
left-handed chiral form, transform as complex representations of the gauge
group.  A chiral gauge theory is defined as being irreducibly chiral if it does
not contain any vectorlike subsector.  In this case, the chiral gauge symmetry
forbids any fermion mass terms in the underlying Lagrangian.  In order for the
theory to be renormalizable, one requires that it must be free of any triangle
anomalies in gauged currents. 

In this paper we shall analyze a variety of chiral gauge theories
with direct-product gauge groups of the form
\beq
G = \bigotimes_{i=1}^{N_G} G_i 
\label{ggen}
\eeq
with fermion contents chosen so that all non-Abelian gauge interactions are
asymptotically free.  The reason for this choice is that this enables one to
carry out perturbative calculations at a sufficiently large Euclidean
energy/momentum scale, $\mu$, in the deep ultraviolet (UV).  As the theory
evolves from the UV to the infrared (IR), these non-Abelian gauge interactions
thus grow in strength.  We restrict here to theories without fundamental scalar
fields. The gauge group $G$ is taken to contain $N_{NA}$ non-Abelian factor
groups, and, by convention, we order the factor groups in the tensor product
(\ref{ggen}) so that these non-Abelian factor groups come before any possible
Abelian factor group(s).  

The main question that we investigate is how patterns of dynamical gauge
symmetry breaking depend on the structure of the direct product gauge group
(\ref{ggen}) and on the relative strengths of the gauge couplings for various
factor groups $G_i \subset G$ that become strong in the IR. We assume
that if $G$ contains any Abelian gauge interaction, it is weakly coupled at
high scales $\mu$ in the UV; given that such a gauge interaction has a positive
beta function, this implies that the Abelian coupling will also remain weak at
lower scales in the infrared.  Our study of a variety of direct-product chiral
gauge theories shows how the patterns of symmetry breaking depend on the
structure of $G$ and on the relative sizes of the gauge couplings corresponding
to factor groups in the direct product.  If the gauge coupling for one of these
factor groups $G_i \subset G$ gets sufficiently strong and dominates over the
other(s), then it can produce bilinear fermion condensates that can self-break
the $G_i$ symmetry itself and/or break other gauge symmetries $G_j \subset G$.

An example of this dependence of the type of gauge symmetry breaking upon the
relative strengths of gauge couplings in a direct-product chiral gauge theory
is provided by a modification of the Standard Model (SM) with the same $N_G=3$
gauge group $G_{SM} = {\rm SU}(3)_c \otimes {\rm SU}(2)_L \otimes {\rm U}(1)_Y$
and with the usual fermion content, but with the Higgs field removed. 
If, at a given scale $\Lambda_{QCD}$, the color SU(3)$_c$ gauge
coupling becomes sufficiently large while the SU(2)$_L$ (and U(1)$_Y$) gauge
couplings are weak, then the SU(3)$_c$ gauge interaction produces a bilinear
quark condensate $\langle \bar q q \rangle$, which dynamically breaks the
electroweak gauge symmetry $G_{EW} = {\rm SU}(2)_L \otimes {\rm U}(1)_Y$ to
electromagnetic U(1)$_{em}$, giving masses to the $W$ and $Z$ bosons.  
Indeed, this was a motivation for models of dynamical electroweak symmetry
breaking by a hypothesized vectorial, asymptotically free gauge interaction
that would become strongly coupled at the TeV scale and would produce bilinear
fermion condensates involving a set of fermions that are nonsinglets under
$G_{EW}$ \cite{tc}.  In this scenario, as well as in quantum chromodynamics
(QCD) itself, the interaction that becomes strong is vectorial and breaks a
weakly coupled chiral gauge interaction to a vectorial subgroup gauge symmetry,
namely U(1)$_{em}$.  In contrast, as discussed in \cite{smr} in the context of
the gedanken SM theory with no Higgs field, if the SU(2)$_L$ gauge coupling
were sufficiently large at a given reference scale, while the SU(3)$_c$ gauge
coupling were weak, then a very different pattern of symmetry breaking would
occur: this SU(2)$_L$ gauge interaction would produce bilinear
fermion condensates that preserve the SU(2)$_L$ gauge invariance but 
break SU(3)$_c$ to SU(2)$_c$, and break U(1)$_Y$, giving masses to the
gluons in the coset ${\rm SU}(3)_c/{\rm SU}(2)_c$ and to the hypercharge
gauge boson. 

Chiral gauge theories (without scalars) that are asymptotically free and can
therefore become strongly coupled at low energies have been of interest in the
past for several reasons. One motivation involved an effort to understand the
pattern of quark and lepton generations. Since the respective lower bounds on
the compositeness scales of these Standard-Model fermions are much larger than
their masses, a plausible approach was to begin by using a theoretical
framework in which they were massless.  Strongly coupled irreducibly chiral
gauge theories are a natural candidate for such a framework, since the chiral
gauge invariance forbids any fermion mass terms. If such a theory satisfies the
't Hooft global anomaly-matching conditions, then, as the gauge coupling
becomes sufficiently strong in the infrared, the gauge interaction could
confine and produce massless gauge-singlet composite spin-1/2 fermions
\cite{thooft1979}-\cite{cgt2}.

A different motivation for studying strongly coupled chiral gauge theories
arose in the context of models that sought to explain both dynamical
electroweak symmetry breaking and fermion mass generation.  In terms of
low-energy effective Lagrangians, this involved the above-mentioned new
vectorial gauge interaction that would become strong at the TeV scale and
produce bilinear fermion condensates, in conjunction with a set of four-fermion
operators that could give rise to quark and lepton masses \cite{tc,etc}.  A
next step was the construction of ultraviolet-completions of these theories
that would have the potential to explain not only the Standard-Model fermion
masses in a given generation, but also the existence of a generational
hierarchy of fermion masses.  A basic property of a chiral gauge theory is that
if it becomes strongly coupled, it can produce bilinear fermion condensates
that self-break the gauge symmetry \cite{georgi79,mac}. Reasonably UV-complete
models for dynamical electroweak symmetry breaking and Standard-Model fermion
mass generation made use of this feature (e.g.,
\cite{at94},\cite{nt}-\cite{constraints}). These involved strongly coupled
chiral gauge interactions that led to the formation of various fermion
condensates which broke the initial chiral gauge symmetry in a sequence of
stages that might plausibly explain the SM fermion masses and their
generational hierarchy.  This sequential breaking was such as to yield, as a
residual symmetry, the vectorial gauge symmetry that is strongly coupled at the
TeV scale.  Ref. \cite{nt} used a direct-product chiral gauge group with two
strongly coupled gauge interactions and pointed out that different patterns of
sequential gauge symmetry breaking (denoted $G_a$ and $G_b$ in \cite{nt}) could
occur, depending on the relative sizes of gauge couplings corresponding to
these two factor groups.  A similar phenomenon was noted in other models
studied in \cite{lrs}. It is this interesting property of the nonperturbative
behavior of direct-product chiral gauge theories that we wish to explore
further here.  

Another motivation for the present study is the fact that
patterns of gauge symmetry breaking by Higgs fields depend on parameters in the
Higgs potential $V$, which one can choose at will, subject to the constraint 
that $V$ should be bounded from below. In contrast, once one has specified the
gauge and fermion content of a chiral gauge theory, together with the values of
the gauge couplings at a reference point (which is naturally chosen to be in
the deep UV for theories with asymptotically free non-Abelian gauge
interactions), then the dynamics determines the pattern of gauge symmetry
breaking uniquely \cite{isb}.

This paper is organized as follows.  In Section \ref{methods_section} we 
discuss our general theoretical framework, methods of analysis, and a
classification of direct-product chiral gauge theories.  Section 
\ref{construction_section} contains some useful procedures for the 
construction of (anomaly-free, asymptotically free) chiral gauge theories.  
In Sections \ref{gcavxsu2_section}-\ref{sunxsum_section} we 
study a variety of different chiral gauge theories with a direct-product 
gauge groups and fermion contents. These involve both unitary and orthogonal
gauge groups and elucidate how the patterns of dynamical symmetry breaking
depend on the structures of the respective theories. 
Our conclusions are contained in Section \ref{conclusion_section}.

% ===========================================================================

\section{Classification of Groups and Methods of Analysis} 
\label{methods_section}

In order to explore the nonperturbative behavior of direct-product chiral gauge
theories, it is useful to have a general classification of these theories and
general methods for analyzing them.  We discuss these in this section. As
stated above, we consider direct-product chiral gauge theories with gauge
groups of the form (\ref{ggen}) with fermion content $\{f \}$ chosen such that
the theory is free of any anomalies in gauged currents and free of any global
SU(2) Witten anomalies, and also such that all non-Abelian gauge interactions
are asymptotically free.  Unless otherwise indicated, we will, with no
loss of generality, write all fermions as left-handed chiral components.

To describe our classification system, we first introduce some notation. We
generically denote a group that has only real or pseudoreal representations as
$G_r$ and a group that has complex representations as $G_c$.  A group $G_r$
cannot, by self, be the gauge group of a chiral gauge theory, although it can
appear as a factor group in a chiral gauge theory.  A group $G_r$ has zero
anomaly, while, in general, a group $G_c$ has nonzero anomalies $A_{\cal R}$
for its representations (see Eq. (\ref{anomdef})), which we will indicate by
the symbol $G_{ca}$.  If a group $G_c$ has no anomaly, i.e., $A_{\cal R}=0$ for
all ${\cal R}$, then it is commonly termed ``safe'' ($s$) \cite{anomalyfree},
and we denote it as $G_{cs}$. Of course, a group $G_r$ is automatically
safe. Thus, the generic class $G_s$ includes $G_r$ and $G_{cs}$.

We may then classify a chiral gauge theory with the direct-product gauge group
({\ref{ggen}) by an $N_G$-dimensional vector indicating the nature of the
  factor groups involved in the direct product.  If $N_G=1$, there are two
  possibilities: (i) $(ca)$, e.g., SU($N$) with $N \ge 3$, and (ii) $(cs)$,
  e.g., SO($4k+2$) for $k \ge 2$ or the exceptional group E$_6$ 
  \cite{anomalyfree,groupinv,rsv99}.   For $N_G=2$, the possibilities are
\beq 
N_G=2: \quad (ca,r), \ (cs,r), \ (ca,ca), \ (ca,cs), \ (cs,cs) \ , 
\label{ng2gtypes}
\eeq
where we do not distinguish the order of factor groups, so, for example, 
$(cs,ca)$ and $(ca,cs)$ are the same type. 

Let us consider a factor group $G_i$ in (\ref{ggen}) which is of the form
$G_{ca}$, and set the gauge couplings of the other factor groups to zero. If
the resultant $G_{ca}$ theory is vectorial ($v$), then we denote this as
$G_{cav}$. This is the case, for example, with the color SU(3)$_c$ factor group
in the Standard Model. Thus, a further classification of direct-product chiral
gauge theories can be carried out in which, for for each factor group of the
form $G_{ca}$, one distinguishes whether or not it is of the form $G_{cav}$.
The Standard Model gauge group is of the type $(cav,r,ca)$ in this
classification.  We illustrate the classification of some chiral gauge theories
considered in this paper in Table \ref{cgt_types}.

\begin{table}
\caption{\footnotesize{Classification of some direct-product chiral gauge 
theories.  See text for further discussion.}}
\begin{center}
\begin{tabular}{|c|c|c|} \hline\hline
Type    & $N_G$ &   $G$ \\
\hline
$(ca,r)$, $(cav,r)$ &2 & ${\rm  SU}(N) \otimes {\rm SU}(2)$ with $N \ge 3$ \\
$(cav,cav)$    &2 & ${\rm SU}(N) \otimes {\rm SU}(M)$ with $N,M \ge 3$  \\
$(r,ca)$    & 2 & ${\rm SU}(2) \otimes {\rm U}(1)$ \\
$(ca,ca)$, $(cav,ca)$ & 2 & ${\rm SU}(N) \otimes {\rm U}(1)$ with $N \ge 3$ \\ 
$(cav,r,ca)$&3 & ${\rm SU}(N_c) \otimes {\rm SU}(2)_L \otimes {\rm U}(1)_Y$\\
$(cav,r,r)$ &3 & ${\rm SU}(N) \otimes {\rm SU}(2)_L \otimes {\rm SU}(2)_R$ \\
$(cav,r,r,cav)$&4&
${\rm SU}(N_c)\otimes{\rm SU}(2)_L\otimes{\rm SU}(2)_R\otimes {\rm
  U}(1)_{B-L}$\\
$(cs,r)$    &2 & ${\rm SO}(4k+2) \otimes {\rm SU}(2)$ with $k \ge 2$  \\
$(cs,cav)$   &2 & ${\rm SO}(4k+2) \otimes {\rm SU}(N)$ with $N \ge 3$ \\
$(cs,cs)$   &2 & ${\rm SO}(4k+2) \otimes {\rm SO}(4k'+2)$ with $k,k' \ge 2$ \\
\hline\hline
\end{tabular}
\end{center}
\label{cgt_types}
\end{table}

Our requirement that each non-Abelian factor group in the direct product
(\ref{ggen}) is asymptotically free enables us to describe the theory
perturbatively in the deep ultraviolet.  We discuss the evolution from the UV
to the IR next.  To each factor group $G_i$, $i=1,...,N_G$,
there corresponds a running gauge coupling $g_i(\mu)$, and we define
$\alpha_i(\mu) = g_i(\mu)^2/(4\pi)$ and $a_i(\mu) \equiv g_i(\mu)^2/(16\pi^2)$.
The argument $\mu$ will often be suppressed in the notation.
The UV to IR evolution of the gauge coupling is determined by the 
beta function, $\beta_{g_i} = dg_i/dt$, or equivalently, 
$\beta_{G_i} = d\alpha_i/dt = [g/(2\pi)]\beta_{g_i}$, 
where $dt = d\ln \mu$. This has the series expansion 
\beqs
\beta_{G_i} & = &-8\pi a_i^2 \bigg [ b_{G_i,1\ell} + 
\sum_{j=1}^{N_G} b_{G_i,2\ell;ij} a_j \cr\cr
& + & \sum_{j,k=1}^{N_G} b_{G_i,3\ell;ijk}a_ja_k + ... \bigg ] \ , 
\label{beta}
\eeqs
where an overall minus sign is extracted and the dots $...$ indicate 
higher-loop terms.  Here, $b_{G_i,1\ell}$ is the one-loop (denoted $(1\ell$)) 
coefficient, multiplying 
$a_i^2$, $b_{G_i,2\ell;ij}$ is the two-loop coefficient, multiplying 
$a_i^2a_j$, and so forth for higher-loop loops.  
The property of asymptotic freedom for the non-Abelian gauge interactions means
that $\beta_{G_i} < 0$ for small $\alpha_i$, $i=1,...,N_{NA}$.  The set
(\ref{beta}) constitutes a set of $N_G$ coupled nonlinear first-order 
ordinary differential equations for the quantities $\alpha_i$, $i=1,...,N_G$.
To leading order, i.e., to one-loop order, the set of
differential equations decouple from each, and one has the simple solution for
each $i \in \{1,...,N_G\}$:
\beq
\alpha_i(\mu_1)^{-1} = \alpha_i(\mu_2)^{-1} - \frac{b_{G_i,1\ell}}{2\pi} \, 
\ln\Big ( \frac{\mu_2}{\mu_1} \Big ) \ , 
\label{alfsol}
\eeq
where we take $\mu_1 < \mu_2$.  

In the following discussion, we assume that the fundamental Lagrangian has no
fermion mass terms, so that all fermion masses are generated dynamically by
chiral symmetry breaking. For a pair of gauge interactions corresponding to the
factor groups $G_i$ and $G_j$ in Eq. (\ref{ggen}), the respective beta
functions $\beta_{G_i}$ and $\beta_{G_j}$ in the deep UV are fixed once we
choose the fermion content of a given theory.  The values of the corresponding
$\alpha_i(\mu_1)$ and $\alpha_j(\mu_1)$ at lower Euclidean scales are
determined by (i) the initial values of $\alpha_i(\mu_2)$ and $\alpha_j(\mu_2)$
in the UV; (ii) the values of $\beta_{G_i}$ and $\beta_{G_j}$; and (iii) the
occurrence of bilinear fermion condensate formation at some scale(s) as the
theory evolves from the deep UV toward the IR, which produce dynamical masses
for the fermions involved in these condensates.  Since we do not assume that
the direct-product group (\ref{ggen}) is contained in a simple group in the
deep UV, we are free to consider various different orderings of the sizes of
the couplings $\alpha_i(\mu_2)$ in the UV.  Furthermore, because of the
condensation process(es) (iii), the fermions involved in these condensates,
together with gauge bosons corresponding to broken generators of gauge
symmetries, acquire dynamical masses and are integrated out of the low-energy
effective field theories that are applicable as the Euclidean reference scale
decreaes below each condensation scale.  The reduction in massless particle
content in (iii) produces changes in the beta functions of the gauge
interactions involved.  Because of this, even if $\beta_{G_i} > \beta_{G_j}$
with all fermions initially present in the deep UV, it can happen that at a
lower scale this inequality is reversed.  The variation of gauge couplings in
the deep UV embodied in the input (i) above was carried out in the earlier work
\cite{nt} where both of the cases of relative sizes of $\alpha_{ETC}$ and
$\alpha_{HC}$ were considered, and in \cite{smr}, where both of the cases of
relative sizes of couplings for SU(3)$_c$ and SU(2)$_L$ were considered.
Henceforth, for notational simplicity, we set $b_{G_i,1\ell} \equiv b_{G_i,1}$.
There have been a number of interesting studies of renormalization-group (RG)
flows in quantum field theories with multiple interaction couplings using
perturbatively calculated beta functions, e.g., \cite{rgflows}.  Here, as in
the earlier works involving gauge theories with multiple gauge couplings
\cite{nt,ckm,smr}, we will focus on the nonperturbative phenomenon of fermion
condensate formation and the associated pattern of gauge symmetry breaking. The
one-loop result (\ref{alfsol}) will be sufficient for our purposes here since
we focus on this nonperturbative fermion condensate formation.  These
condensates also generically break global chiral symmetries.

In general, a fermion condensate may involve different fermion fields or
the same fermion field.  If the fields are the same, we may write the bilinear
fermion operator product abstractly as follows.  Assume that the gauge group
$G$ in Eq. (\ref{ggen}) contains $t \le N_G$ non-Abelian factors $G_k$ and that
the relevant fermion field $f$ transforms as the representation ${\cal R}
\equiv ({\cal R}_1,...,{\cal R}_t)$ under the direct product of these
non-Abelian factor groups.  Then the bilinear fermion product of a given
fermion field is
\beq
f^T_{{\cal R},i,L} C f_{{\cal R},j,L} \ , 
\label{ff}
\eeq
where $C$ is the Dirac conjugation matrix, gauge group indices are suppressed
in the noation, and $i,j$ are copy (flavor) indices. 
From the property $C^T = -C$ together with the anticommutativity
of fermion fields, it follows that the bilinear fermion operator product 
(\ref{ff}) is symmetric under interchange of the order of fermion
fields and therefore is symmetric in the overall product 
\beq
\Big [ \prod_{k=1}^t ({\cal R}_k \times {\cal R}_k) \Big ] \, S_{ij} \ , 
\label{symproperty}
\eeq
where $S_{ij}$ abstractly denotes the symmetry property under interchange of
flavors, with $S_{ij}=(ij)$ and $S_{ij}=[ij]$ for symmetric and antisymmetric
flavor structure, respectively.  For example, for the case $t=N_g=2$ and flavor
indices $i,j$, the symmetry property (\ref{symproperty}) means that $f^T_{i,L}
C f_{j,L}$ is of the form $(s,s,s)$, $(s,a,a)$, $(a,s,a)$, or $(a,a,s)$, where
here $s$ and $a$ indicate symmetric and antisymmetric and the three entries
refer to the representations ${\cal R}_1$ of $G_1$, ${\cal R}_2$ of $G_2$, and
$S_{ij}$.  Thus, as an illustration, in the last case, $(a,a,s)$, the product
(\ref{ff}) would transform as antisymmetric representations in the
Clebsch-Gordan products of ${\cal R}_j \times {\cal R}_j$ for $j=1,2$ and would
be symmetric in flavor indices, with $S_{ij}=(ij)$, and so forth for other
cases.

The main perturbative information that we will use is the 
one-loop coefficients of the beta functions for the non-Abelian gauge 
interactions. We require that these interactions must be asymptotically free so
that we have perturbative control over them in the deep UV. 
If $\alpha_i(\mu)$ becomes strong, i.e., O(1) in the IR, one can no longer use
perturbative methods reliably, but one can make use of several approximate
methods to explore possible nonperturbative properties of the theory. First,
one may investigate whether the fermions in the theory satisfy the 't
Hooft anomaly-matching conditions.  For this purpose, one determines the global
flavor symmetry group of the theory is invariant and then checks whether 
candidate operators for gauge-singlet composite spin-1/2
fermions match the anomalies in the global flavor
symmetries. If this necessary condition is satisfied, then it is possible 
that in the infrared the strong chiral gauge interaction could confine and
produce massless composite spin-1/2 fermions. 

A different possibility in a strongly coupled chiral gauge theory is that the
gauge interaction can produce bilinear fermion condensates.  This will be the
main focus of our analysis here.  In an irreducibly chiral theory these
condensates break one or more gauge symmetries, as well as global flavor
symmetries.  A commonly used method for suggesting which type of condensate is
most likely to form in this case is the most-attractive-channel (MAC) method
\cite{mac}.  For possible condensation of chiral fermions in the
representations ${\cal R}_{G_i,1}$ and ${\cal R}_{G_i,2}$ of the factor group
$G_i$ in (\ref{ggen}) in various channels of the form 
${\cal R}_{G_i,1} \times {\cal R}_{G_i,2} \to {\cal R}_{G_i,cond.}$, 
the MAC approach predicts that the condensation
will occur in the channel with the largest (positive) value of the quantity
\beq
\Delta C_2 \equiv C_2({\cal R}_{G_i,1}) + C_2({\cal R}_{G_i,2})
-C_2({\cal R}_{G_i,cond.}) \ ,
\label{deltac2}
\eeq
where $C_2({\cal R})$ is the quadratic Casimir invariant for the 
representation ${\cal R}$
(see Appendix \ref{group_invariants}).  This is only a rough measure, based on
one-gluon exchange.  The form of the condensate determines the resultant
symmetry and form of vacuum alignment \cite{vacalign}.

% =======================================================================

\section{Methods for Constructing Chiral Gauge Theories}
\label{construction_section}

In this section we mention some useful methods for constructing 
anomaly-free direct-product chiral gauge theories.  

% =======================================================================

\subsection{Reduction Method}
\label{reduction_section}

Let us say that we have a chiral gauge theory with the $N_G$-fold direct
product gauge group (\ref{ggen}) and a given fermion content that satisfies the
constraints that the theory must be free of any anomaly in gauged currents, any
possible global SU(2) anomaly, and, if $G$ includes abelian factor groups,
also any mixed gravitational-gauge anomaly.  One can then construct a set of
chiral gauge theories by a process of reduction, setting one or more of the
gauge couplings $\{g_1,...,g_{N_G} \}$ equal to zero.  As an example, if one
starts with a modified and extended Standard Model with gauge group (\ref{gsm})
and fermion content (\ref{ql_gsm})-(\ref{ll_gsm}) below, of type $(cav,r,ca)$,
then (i) by turning off the ${\rm SU}(N_c)$ gauge coupling, one gets an ${\rm
  SU}(2)_L \otimes {\rm U}(1)_Y$ gauge theory of type $(r,ca)$; (ii) by turning
off the SU(2)$_L$ gauge coupling, one gets an ${\rm SU}(N_c) \otimes {\rm
  U}(1)_Y$ gauge theory of type $(cav,ca)$; and (iii) by turning off the
U(1)$_Y$ coupling, one gets an ${\rm SU}(N_c) \otimes {\rm SU}(2)_L$ gauge
theory of type $(cav,r)$.  Given that the original theory has the requisite
property that all non-Abelian gauge interactions are asymptotically free, the
theory derived by turning off some gauge coupling(s) also has this property.

% ======================================================================

\subsection{Extension Method to Construct $G = \tilde G \otimes G_s$ Theories} 
\label{gcaxgs_section}

Here we present a method for constructing a direct-product chiral gauge theory
with an $(N_G+1)$-fold direct-product gauge group, starting from a given chiral
gauge theory with an $N_G$-fold direct-product gauge group $\tilde G$ by
adjoining a safe group $G_s$ to $\tilde G$ to produce 
\beq
G = \tilde G \otimes G_s
\label{gtildegs}
\eeq
and extending the fermion
representations of $\tilde G$ to those of $G = \tilde G \otimes G_s$. 
Here $G_s$ may be $G_r$ or $G_{cs}$. The procedure is as follows:

\begin{enumerate}

\item Start with an anomaly-free chiral gauge theory with the $N_G$-fold
  gauge group $\tilde G=\bigotimes_{i=1}^{N_G} G_i$ and a set of fermion 
  representations $\{{\cal R}_{\tilde G} \}$,  where each of these is 
\beq
{\cal R}_{\tilde G} = ({\cal R}_{G_1},...,{\cal R}_{G_{N_G}})
\label{repgtilde}
\eeq

\item Choose the safe group $G_s$, of type $G_r$ or $G_{cs}$, i.e., either a
  group with real representations, such as SU(2), or a safe group with 
  complex representations, such as SO($4k+2$) with $k \ge 2$ or the exceptional
  group E$_6$.

\item Extend each fermion representation ${\cal R}_{\tilde G}$ of $\tilde G$ to
  a representation ${\cal R}_G$ of $G$ using a single representation 
  ${\cal R}_{G_s}$ of $G_s$ to form 
  ${\cal R}_G = ({\cal R}_{\tilde G}, {\cal R}_{G_s})$. 
  As far as the $\tilde G$ group is 
  concerned, this simply amounts to a replication of its original
  (anomaly-free) fermion content by ${\rm dim}({\cal R}_{G_s})$ copies, so the
  resulting extended fermion content is also anomaly-free.  

\item Apply the constraint that if the safe group is $G_s={\rm SU}(2)$, then
  the resultant theory must be free of a
  global SU(2) Witten anomaly associated with the homotopy group
$\pi_4({\rm SU}(2))={\mathbb Z}_2$ \cite{wittensu2,homotopy}. 
  With ${\cal R}_{G_s}=\fund$, 
  the necessary and sufficient condition to satisfy this constraint is that 
  the total number of SU(2) doublets is even \cite{wittensu2}. 

\item Apply the constraints that each of the gauge interactions corresponding
  to non-Abelian factor groups in $\tilde G$ must remain asymptotically
  free in the larger group $G$, and the $G_s$ gauge interaction must also be 
  asymptotically free.

\end{enumerate}

This method can be used to construct many types of direct-product chiral gauge
groups.  Among the $N_G=2$ cases, for example, these types include 
all of the ones listed in Eq. (\ref{ng2gtypes}).

% ========================================================================

\section{$G_{cav} \otimes {\rm SU}(2)$ Theories}
\label{gcavxsu2_section}

In this section we construct and study a class of
$N_G=2$ direct-product chiral gauge theories with a gauge group
\beq
G_1 \otimes G_2 = G_{cav} \otimes {\rm SU}(2) \ . 
\label{gcavxsu2}
\eeq
This class is the special case $(cav,r)$ of the class $G_{ca} \otimes G_r$
discussed in Section \ref{methods_section} in 
which $G_{ca}=G_{cav}$, i.e., $G_{ca}$ is a group with complex
representations and ${\cal A}_{\cal R} \ne 0$ and the fermion content is such
that if the SU(2) gauge interaction is turned off, then the $G_{cav}$ gauge
interaction is vectorial. This property guarantees that
there is no cubic triangle anomaly in gauged currents in the $G_{cav}$
sector. Furthermore, as already indicated above, since SU(2) has
(pseudo)real representations, it has no anomaly. 
The only anomaly constraint is then the requirement that the SU(2) group must
be free of a global anomaly. We consider theories of this type with chiral
fermion content (written here as left-handed)
\beq
\{ f_{ns,ns} \} = \sum_{\cal R} p_{_{\cal R}} \, ({\cal R},\fund) \ , 
\label{f_nsns_gcavxsu2}
\eeq
\beq
\{ f_{ns,s} \} = 2 \sum_{\cal R} p_{_{\bar{\cal R}}} \, 
(\bar{{\cal R}},1) \ , 
\label{f_nss_gcavxsu2}
\eeq
and optionally, 
\beq
\{ f_{s,ns} \} = p_1 \, (1,\fund) \ , 
\label{f_sns_gcavxsu2}
\eeq
where the subscripts $ns$ and $s$ are abbreviations for ``nonsinglet'' and
``singlet''; ${\cal R}$ denotes a (nonsinglet) representation of the group
$G_1$; and the first and second entries in subscripts and in the parentheses
refer to the representations of $G_{cav}$ and SU(2)$_L$, respectively, with
$\fund$ being the fundamental representation in standard Young tableaux
notation. 

If the fermion sector includes only a single ${\cal R}$, then we set
$p_{\cal R} \equiv p$ for brevity. We shall use interchangeably a notation with
Young tableaux and dimensionalities to identify the representation: 
$({\cal R},\fund) \leftrightarrow ({\rm dim}({\cal R}),2)$. In general, we will
allow for several types of (nonsinglet) representations ${\cal R}$, but will
focus on minimal theories with only one ${\cal R}$.  The subscript indices 
$i,j$ are copy (``flavor'') indices, and the total number of copies of the
$f_{ns,ns}$ fermions transforming as the ${\cal R}$ representation of $G_1$
is denoted $p_{\cal R}$. We shall mainly focus on irreducibly chiral theories,
i.e., those for which the chiral gauge theory forbids any bare mass terms, but
we shall also discuss some chiral gauge theories with vectorlike subsectors. 
The global symmetries depend on $p$ and $p_1$; we will discuss them
for specific models below. 

The number of SU(2) chiral fermion doublets in this theory, which we shall
denote $N_d$, is
\beq
N_d = p_1+ \sum_{\cal R} \, p_{_{\cal R}} \, {\rm dim}(\cal R) \ . 
\label{nd}
\eeq
The condition that the SU(2) gauge sector must be 
free of a global anomaly is that 
\beq
N_d \ \ {\rm is \ \ even} \ . 
\label{nd_even}
\eeq
Because $N_d$ is necessarily even, one could take half of the left-handed 
SU(2)-doublet fermions, rewrite them as right-handed charge-conjugates, and
thereby put the SU(2) gauge interaction into vectorial form. 

As noted, we shall also impose two further requirements on the theory, namely
that the $G_1$ and the SU(2) gauge interactions must both be asymptotically
free.  From the general results in \cite{b1}, we find that the one-loop 
coefficient of the beta function of the $G_1$ gauge interaction is
\beq
b_{1,G_1}=\frac{1}{3}\Big [11C_2(G_1)- 8\sum_{\cal R}p_{_{\cal R}} \, 
T({\cal R}) \Big ] \ , 
\label{b1_g1_g1xsu2}
\eeq
so the requirement that the $G_1$ gauge interaction should be
asymptotically free implies that 
\beq 
\sum_{\cal R} p_{_{\cal R}} \, T({\cal R})  < \frac{11C_2(G_1)}{8} \ .  
\label{pbound_g1xsu2_g1_af}
\eeq
Here and below, if $p_1=0$ and the theory contains fermions in one (nonsinglet)
representation ${\cal R}$ of $G_1$, then only nonzero values of $p_{\cal R}
\equiv p$ are relevant, since if $p=0$, then the theory is a pure
(direct-product) gauge theory and hence is not a chiral gauge theory.

The one-loop coefficient of the beta function of the SU(2) gauge interaction is
\beqs
& & b_{1,{\rm SU}(2)_L} = \frac{1}{3}(22-N_d) \cr\cr
& = &  \frac{1}{3}\Big [ 22- \Big (p_1 + \sum_{\cal R} p_{_{\cal R}} \, {\rm
    dim}(\cal R) \Big ) \Big ] \ , 
\label{b1_su2_g1xsu2}
\eeqs
so the requirement that the SU(2) gauge interaction should be 
asymptotically free implies that 
\beq 
p_1 + \sum_{\cal R} p_{_{\cal R}} \, {\rm dim}({\cal R})  < 22 \ .  
\label{pbound_g1xsu2_su2_af}
\eeq
%

% =======================================================================

\section{${\rm SU}(N) \otimes {\rm SU}(2)$ Theories}
\label{sunxsu2_section}

In this section we construct and study several models with a direct-product
gauge group of the form (\ref{gcavxsu2}) with the first gauge group being 
SU($N$), i.e., with
\beq
G = G_1 \otimes G_2 = {\rm SU}(N) \otimes {\rm SU}(2)
\label{gsunxsu2}
\eeq
and various chiral fermion contents, which we denote as Models A, B, and C.
All three of these models are of type $(cav,r)$, as indicated in Table
\ref{cgt_types}.

% =======================================================================

\subsection{Model A}
\label{sunxsu2a_section}

The first model that we consider, denoted Model A, is a minimal one in three
respects: (i) it contains no $G_1$-singlet fermions, i.e., $p_1=0$; (ii) the
fermions transform according to only one representation ${\cal R}$ of $G_1$ and
its conjugate; and (iii) this representation ${\cal R}$ is the simplest
nontrivial one, namely the fundamental, ${\cal R}=\fund$.  The chiral fermions
are
\beq
\psi^{a,\alpha}_{i,L}, \ i=1,...,p \ : \quad p \, (\fund,\fund) = p \, (N,2), 
\label{ql_sunxsu2a}
\eeq
and
\beq
\chi_{a,j,L}, \ j=1,...,2p \ : \quad  2p \, (\overline{\fund},1) = 2p \, 
(\bar{N},1) \ . 
\label{qcl_sunxsu2a}
\eeq
Here, $a$ and $\alpha$ are SU($N$) and SU(2) gauge indices and $i, j$
are copy (``flavor'') indices.  For $N \ge 3$, the chiral gauge symmetry
forbids any bare mass terms for the fermions.  In contrast, if $N=2$, then 
gauge-invariant bare mass terms such as 
\beq
\epsilon^{ab} \chi_{a,i,L}^T C \chi_{b,j,L} \ , \quad i \ne j, \quad 
1 \le i, j \le 2p
\label{chichi_su2su2}
\eeq
and
\beq
\epsilon_{ab}\epsilon_{\alpha\beta} 
\psi^{a,\alpha \ T}_{i,L} C \psi^{b,\beta}_{j,L} \ , \quad 
1 \le i, j \le p
\label{zetazeta_su2su2}
\eeq
can occur. Closely related to this, if $N=2$, then the SU($N$) and SU(2)
gauge interactions can both be written in vectorial form, so the theory is not
a chiral gauge theory.  Therefore, henceforth we shall assume that $N \ge 3$
for this class of theories.  In the notation introduced above, 
the fermion content of this Model A can be categorized as being of the form 
\beq
\{ f_{ns,ns}, \ f_{ns,s} \} \ . 
\label{fform_sunxsu2a}
\eeq
The fermion terms in the Lagrangian for this model are
\beq
{\cal L} = \sum_{j=1}^p \bar\psi_{j,L} i\Dslash \, \psi_{j,L} + 
\sum_{j=1}^{2p} \bar \chi_{j,L} i\Dslash \, \chi_{j,L} \ , 
\label{lagrangian_sunxsu2a}
\eeq
(where we have indicated the sums over flavor indices explicitly). 
In connection with the discussions in Sections \ref{reduction_section} and
\ref{gsm_section}, we note that one realization of a Model A theory is the
gauge and quark sector of the generalized Standard Model with the Higgs field
removed, the weak hypercharge gauge coupling turned off, and with the
identifications $N=N_c$ and $p=N_g$, where $N_g$ denotes the number
of fermion generations. In this case, the correspondence of fermion fields here
and in Eqs. (\ref{ql_gsm}) and (\ref{qr_gsm}) is as given below in
Eqs. (\ref{qell_psi_corrspondence}) and (\ref{qr_chi_correspondence}).  This
correspondence motivates the property that the Lagrangian
(\ref{lagrangian_sunxsu2a}) is diagonal in copy indices; if one were to include
terms of the form $\bar \chi_{j,L} i\Dslash \, \chi_{k,L}$ with $j \ne k$,
some of these would correspond, in the generalized SM, to terms of the form
$\bar u_{j',L} i\Dslash \, d_{k',L}$ that would violate U(1)$_Y$ and
electromagnetic U(1)$_{em}$ gauge symmetries.  Although Model A has no U(1)$_Y$
factor, we will restrict the Lagrangian to the form (\ref{lagrangian_sunxsu2a})
which could be derived from the generalized SM by the reduction process of
Section \ref{reduction_section}.

For this Model A, the condition that the SU(2)
gauge sector should be free of a global anomaly is 
\beq
N_d = pN \ {\rm is \ even},  
\label{nopi4anom_sunxsu2a}
\eeq
and we require that this condition must be satisfied.

From the general result (\ref{b1_g1_g1xsu2}), we have, for the one-loop
coefficient of the SU($N$) beta function, 
\beq
b_{1,{\rm SU}(N)} = \frac{1}{3}(11N-4p) \ . 
\label{b1_sun_sunxsu2a}
\eeq
Therefore, the requirement that the SU($N$) gauge interaction should be
asymptotically free, expressed by the inequality (\ref{pbound_g1xsu2_g1_af}), 
reads 
\beq 
p < \frac{11N}{4} \ .  
\label{pbound_sunxsu2a_sun_af}
\eeq

From the general result (\ref{b1_su2_g1xsu2}), we find, for 
the one-loop coefficient of the SU(2) beta function, 
\beq
b_{1,{\rm SU}(2)_L} = \frac{1}{3}(22-pN) \ . 
\label{b1_su2_sunxsu2a}
\eeq
Hence, the requirement that the SU(2) gauge interaction should be
asymptotically free, given by the inequality (\ref{pbound_g1xsu2_su2_af}), is
\beq
p \, N < 22 \ . 
\label{pbound_sunxsu2a_su2_af}
\eeq
In Fig. \ref{allowed_region_sunxsu2a} we show the
boundaries of the region in the $(N,p)$ plane satisfying the inequalities
(\ref{pbound_sunxsu2a_sun_af}) and (\ref{pbound_sunxsu2a_su2_af}). 
The allowed values of $N$ and $p$ are thus the 
integers $N \ge 3$ and $p \ge 1$ in this allowed region that satisfy
the conditions (\ref{pbound_sunxsu2a_sun_af}), 
(\ref{pbound_sunxsu2a_su2_af}), and (\ref{nopi4anom_sunxsu2a}).  
We list these in Table \ref{npvalues_sunxsu2a}. 
\begin{figure}
  \begin{center}
    \includegraphics[height=8cm]{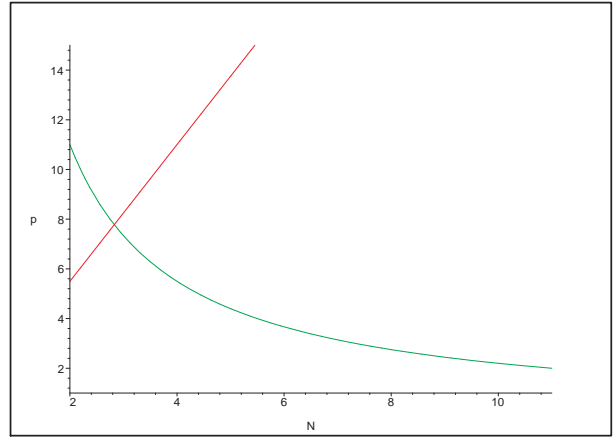}
  \end{center}
\caption{Plot of the region in $N$ and $p$ allowed by the requirement of of
  asymptotic freedom for the SU($N$) and SU(2) gauge interactions in the ${\rm
    SU}(N) \otimes {\rm SU}(2)_L$ Model A chiral gauge theory.  The boundaries
  of this region are given by the line from the inequality
  (\ref{pbound_sunxsu2a_sun_af}) and the hyperbola from the inequality
  (\ref{pbound_sunxsu2a_su2_af}).  The allowed values of $N$ and $p$ are thus
  the integers $N \ge 3$ and $p \ge 1$ in this allowed region that also satisfy
  the condition that the theory must not have any global SU(2) anomaly,
  Eq. (\ref{nopi4anom_sunxsu2a}).  See text for further discussion.}
\label{allowed_region_sunxsu2a}
\end{figure}
\begin{table}
\caption{\footnotesize{Values of $N$ and $p$ in the Model A ${\rm SU}(N)
    \otimes {\rm SU}(2)_L$ chiral gauge theory allowed by the inequalities
    (\ref{pbound_sunxsu2a_sun_af}) and (\ref{pbound_sunxsu2a_su2_af}) arising
    from the constraint of asymptotic freedom for the SU($N$) and SU(2) gauge
    interactions, respectively, and the requirement that the theory must not
    have any global SU(2) anomaly, Eq. (\ref{nopi4anom_sunxsu2a}).  The
    notation $12 \le N_{even} \le 20$ denotes the even values of $N$ in this
    range.  The notation $13 \le N_{odd} \le 21$ denotes the odd values of $N$
    in this range.  For $N \ge 22$, the inequality
    (\ref{pbound_sunxsu2a_su2_af}) has only the trivial solution $p=0$ for
    which the theory is a pure gauge theory with no fermions and hence is not a
    chiral gauge theory.}}
\begin{center}
\begin{tabular}{|c|c|} \hline\hline
$N$ & allowed values of $p$ \\
\hline
 3 &  $p=2, \ 4, \ 6$ \\
 4 &  $1 \le p \le 5$ \\
 5 &  $p=2, \ 4$ \\
 6 &  $1 \le p \le 3$ \\
 7 & $p=2$ \\
 8 & $p=1, \ 2$ \\
 9 & $p=2$ \\
10 & $p=1, \ 2$ \\
11 & no sol. with $p \ne 0$ \\
$12 \le N_{even} \le 20$ & $p=1$ \\
$13 \le N_{odd} \le 21$ & no sol. with $p \ne 0$ \\
$N \ge 22$ & no sol. with $p \ne 0$ \\
\hline\hline
\end{tabular}
\end{center}
\label{npvalues_sunxsu2a}
\end{table}
Several comments are in order concerning these allowed values of $N$ and $p$.
First, as $N$ increases through the value $N=22$, the maximum value of $p$
allowed by the inequality (\ref{pbound_sunxsu2a_su2_af}) decreases below 1, so
that for $N > 22$, this inequality (\ref{pbound_sunxsu2a_su2_af}) has only the
trivial (integral) solution $p=0$ for which the theory is a pure gauge theory
with no fermions and hence not of interest here.  Second, for odd $N$, one sees
that the condition (\ref{nopi4anom_sunxsu2a}) for the theory to be free from a
global SU(2) anomaly restricts $p$ to even values.

We next analyze the UV to IR evolution and gauge symmetry breaking in this
model. If the SU($N$) gauge interaction is sufficiently strong and if it 
dominates over the SU(2) gauge interaction, then this SU($N$) interaction
forms bilinear fermion condensates that break the
SU(2) gauge symmetry.  We denote the scale at which this occurs as $\Lambda$. 
As regards the SU($N$) gauge interaction, the 
most attractive channel for fermion condensation is 
\beq
{\rm SU}(N): \quad 
\fund \times \overline{\fund} \to 1, 
\label{nnbar_to_1_sunxsu2}
\eeq
in terms of Young tableaux, or equivalently, 
$N \times \bar N \to 1$, in terms of the dimensionalities of the
SU($N$) representations, with associated condensates
\beq
\langle \sum_{a=1}^N \, \psi^{a,\alpha \ T}_{i,L} C \chi_{a,j,L} \rangle \ , 
\label{psi_chi_condensate_sunxsu2}
\eeq
where $i \in \{1,...p\}$ and $j \in \{1,...,2p\}$.  (Here and below, when a
condensate is given, it is understood that the hermitian conjugate condensate
is also present.)  This channel has
\beq
\Delta C_2 = 2C_2(\fund)= \frac{N^2-1}{N} \ .
\label{Delta_C2_ffbar_to1}
\eeq
Each of the condensates in Eq. (\ref{psi_chi_condensate_sunxsu2}) breaks the
SU(2) gauge symmetry completely (and is invariant under the SU($N$) gauge
symmetry, as is clear from (\ref{nnbar_to_1_sunxsu2})).  The fermions involved
in these condensates, and the SU(2) gauge bosons, gain dynamical masses of
order $\Lambda$.

If, on the other hand, the SU(2) interaction is sufficiently strong and if
it dominates over the SU($N$) interaction, then this SU(2)
interaction produces bilinear fermion condensates in the most attractive
SU(2) channel $2 \times 2 \to 1$, with associated condensates of the form
\beq
\langle\epsilon_{\alpha\beta}\psi^{a,\alpha \ T}_{i,L} C 
\psi^{b,\beta}_{j,L}\rangle \ . 
\label{psi_psi_condensate_sunxsu2}
\eeq
We denote the scale where this occurs as $\Lambda'$. 
The attractiveness measure for condensate formation in this channel is 
$\Delta C_2 = 2C_2(\fund)=3/2$.  From the general symmetry property
(\ref{symproperty}), it follows that if, as in Eq.
(\ref{psi_psi_condensate_sunxsu2}), one contracts the SU(2) gauge indices
$\alpha$ and $\beta$ antisymmetrically via the SU(2) 
$\epsilon_{\alpha\beta}$ tensor,
then the combination of SU($N$) and generational indices is antisymmetric.
That is, in the operator product (\ref{psi_psi_condensate_sunxsu2}), either the
SU($N$) gauge indices are antisymmetric and the generational indices are
symmetric, so the condensate is proportional to
\beqs
& & \langle\epsilon_{\alpha\beta}(
      \psi^{a,\alpha \ T}_{i,L} C \psi^{b,\beta}_{j,L} 
    - \psi^{b,\alpha \ T}_{i,L} C \psi^{a,\beta}_{j,L} \cr\cr
& +&  \psi^{a,\alpha \ T}_{j,L} C \psi^{b,\beta}_{i,L} 
    - \psi^{b,\alpha \ T}_{j,L} C \psi^{a,\beta}_{i,L})\rangle
\label{psi_psi_condensate_sunxsu2_strongsu2_sunasym}
\eeqs
or the SU($N$) gauge indices are symmetric and the generational indices 
are antisymmetric, so the condensate is proportional to 
\beqs
& & \langle\epsilon_{\alpha\beta}(
      \psi^{a,\alpha \ T}_{i,L} C \psi^{b,\beta}_{j,L} 
    + \psi^{b,\alpha \ T}_{i,L} C \psi^{a,\beta}_{j,L} \cr\cr
& -&  \psi^{a,\alpha \ T}_{j,L} C \psi^{b,\beta}_{i,L} 
    - \psi^{b,\alpha \ T}_{j,L} C \psi^{a,\beta}_{i,L})\rangle \ . 
\label{psi_psi_condensate_sunxsu2_strongsu2_sunsym}
\eeqs
The SU($N$) gauge interaction, although assumed to be weaker than the
SU(2) gauge interaction, is not assumed to be negligible, and it prefers
the condensation channel that is the MAC as regards SU($N$).  Now 
\beq
\Delta C_2 = \frac{N+1}{N} \quad {\rm for} \quad \fund \times \fund \to \asym
\label{Delta_C2_asym_channel}
\eeq
whereas
\beq
\Delta C_2 = -\frac{N-1}{N} \quad {\rm for} \quad \fund \times \fund \to \sym
\label{Delta_C2_sym_channel}
\eeq
so the $\fund \times \fund \to \asym$ channel is the MAC, and indeed, the 
$\fund \times \fund \to \sym$ channel is repulsive.  Therefore, we conclude
that in this case where SU(2) is more strongly coupled than 
SU($N$), the expected condensation channel is, in an obvious notation, 
\beq
(\fund,2) \times (\fund,2) \to (\asym,1)
\label{mac_sunxsu2_strongsu2}
\eeq
with associated condensate
(\ref{psi_psi_condensate_sunxsu2_strongsu2_sunasym}). 
This condensate, which is of the form $\langle T^{[ab]} \rangle$, where 
$T^{[ab]}$ is a rank-2 antisymmetric tensor of SU($N$), breaks SU($N$)
as follows \cite{elias}:
\beq
\langle T^{[ab]}\rangle: \ {\rm SU}(N) \to H = 
 \cases{ {\rm SU}(2) & if $N=3$ \cr
         {\rm SU}(N-2) \otimes {\rm SU}(2) & if $N \ge 4$}
\label{asym_bk_sun}
\eeq
The fermions involved in the condensate and the gauge bosons in the coset 
${\rm SU}(N)/H$ gain dynamical masses of order $\Lambda'$ and are integrated 
out of the low-energy effective field theory that is operative as the reference
scale $\mu$ decreases below $\Lambda'$.  The fermion condensates that form in
both the strong-SU($N$) and strong-SU(2) situations also break global flavor
symmetries.  Since we have already analyzed this sort of global flavor symmetry
breaking in our previous works \cite{cgt,cgt2}, we will not pursue this here,
instead focusing on the gauge symmetry breaking.

% =====================================================================

\subsection{Model B}
\label{sunxsu2b_section}

This model, denoted Model B, has the same gauge group as Model A, but has an
enlarged chiral fermion sector which also contains $p_1 \equiv p'$ copies of
the SU($N$)-singlet, SU(2)-doublet fermion
\beq
\eta^\alpha_{j,L} \ , \  j=1,...,p' \ : \quad p' (1,2) \ . 
\label{eta_eta_sunxsu2b}
\eeq
Thus, the fermion content of Model B can be categorized as being
of the form 
\beq
\{ f_{ns,ns}, \ f_{ns,s}, \ f_{s,ns} \}
\label{fform_sunxsu2b}
\eeq
in the notation of Eq. (\ref{fform_sunxsu2a}).  Depending on the value of $p'$,
these additional fermions may have gauge-invariant bare mass terms of the form
\beq
\epsilon_{\alpha\beta}\eta^{\alpha \ T}_{i,L} C \eta^\beta_{j,L} \ , 
\label{etamassterm}
\eeq
where $i \ne j$ and $1 \le i,j \le p'$.  Using the general symmetry property
(\ref{symproperty}) and taking account of the antisymmetric contraction of the
SU(2) gauge indices $\alpha$ and $\beta$ with the $\epsilon_{\alpha\beta}$
tensor, it follows that the fermion operator in (\ref{etamassterm}) is
automatically antisymmetrized in the flavor indices $i$ and $j$, so if $p'=1$,
then it vanishes identically. If $p' \ge 2$, then the $\{ f_{s,ns} \}$ fermions
constitute a vectorlike subsector in the full chiral gauge theory.

The sector of SU($N$)-nonsinglet fields in Model B is the same as in Model A,
so the SU($N$) gauge interaction is again vectorial and hence is free from any
gauge anomaly, as is the SU(2) gauge interaction.  The condition that the
SU(2) part of the theory should be free of any global anomaly is that the
number of SU(2) doublets, denoted $N_d$, is even, i.e.,
\beq
N_d = pN + p' \ {\rm is \ even}, 
\label{nopi4anom_sunxsu2b}
\eeq
and we require that this condition be satisfied. 

The one-loop coefficient of the SU($N$) beta function, 
$b_{1,{\rm SU}(N)}$, is given by (\ref{b1_sun_sunxsu2a}), as in Model A, so $p$
is subject to the same upper bound from the requirement that the 
SU($N$) interaction must be asymptotically free, namely 
(\ref{pbound_sunxsu2a_sun_af}). 

The one-loop coefficient of the SU(2) beta function is
\beq
b_{1,{\rm SU}(2)_L} = \frac{1}{3}[22-(pN+p')] \ , 
\label{b1_su2_sunxsu2b}
\eeq
so the requirement that the SU(2) gauge interaction should be
asymptotically free implies that
\beq
pN + p' < 22 \ . 
\label{pbound_sunxsu2b_su2_af}
\eeq
The allowed values of $N$, $p$, and $p'$ for Model B are thus the integers $N
\ge 3$, $p \ge 1$, and $p' \ge 1$ satisfying the conditions
(\ref{pbound_sunxsu2a_sun_af}), (\ref{pbound_sunxsu2b_su2_af})), and
(\ref{nopi4anom_sunxsu2b}).  There are too many values to list in a table
analogous to Table \ref{npvalues_sunxsu2a}, but we mention that for $N=3$, the
allowed values of $(p,p')$ are $(1,2k+1)$ with $0 \le k \le 8$; $(2,2k)$ with
$1 \le k \le 7$; $(3,2k+1)$ with $0 \le k \le 5$; $(4,2k)$ with 
$1 \le k \le 4$; $(5,2k+1)$ 
with $0 \le k \le 2$; and the single pair $(6,2)$.  As in Model
A, as $N$ increases, the allowed set of values of $p$ and $p'$ is progressively
reduced, and for sufficiently large $N$, there are no nontrivial solutions to
the three conditions.  For example, for $N=16$, there are only two allowed sets
of $(p,p')$, namely (1,2) and (1,4); for $N=17$, there are again two sets,
namely (1,1) and (1,3), while for $N=18$, there is only one, (1,2), and for
$N=19$, there is only one, (1,1).  For $N \ge 20$, there are no allowed
(nonzero) values of $p$ and $p'$ in this model.

Since Model B is the same as Model A as regards the SU($N$)-nonsinglet fermion
content, it follows that if the SU($N$) gauge interaction is sufficiently 
strong and dominates over the SU(2) interaction, then the resultant 
bilinear fermion condensate formation is the same as in Model A.  

However, if the opposite is the case, i.e., if the SU(2) gauge interaction
is strong enough and dominates over the SU($N$) interaction, then, depending on
the value of $p'$, two additional type of fermion condensates may be produced.
These all have the same SU(2) attractiveness measure, as given before,
namely, $\Delta C_2 = 3/2$ and hence, if SU($N$) interactions are negligible,
they are expected to form at essentially the same Euclidean scale, 
which we again denote as $\Lambda$. Thus, in addition to the condensate(s)
(\ref{psi_psi_condensate_sunxsu2_strongsu2_sunasym}), the SU(2) gauge
interaction can lead to condensation in the channel
\beq 
(1,2) \times (1,2) \to (1,1) 
\label{eta_eta_channel_sunxsu2b}
\eeq
with the associated condensate(s)
\beq
\langle\epsilon_{\alpha\beta}\eta^{\alpha \ T}_{i,L} C \eta^\beta_{j,L}\rangle
\ , 
\label{eta_eta_condensate_sunxsu2b}
\eeq
where $1 \le i,j \le p'$.  From (\ref{symproperty}), it follows that the 
bilinear fermion operator product in 
(\ref{eta_eta_condensate_sunxsu2b}) is antisymmetric in the copy indices 
$i$ and $j$ and hence vanishes identically if $p'=1$.  As is evident from 
(\ref{eta_eta_channel_sunxsu2b}), this condensate
(\ref{eta_eta_condensate_sunxsu2b}) 
preserves the full ${\rm SU}(N) \otimes {\rm SU}(2)_L$ gauge symmetry.  
The fermions
involved in these condensates gain dynamical masses of order the condensation
scale, denoted $\Lambda$, and are integrated out in the low-energy 
effective field theory that is operative as the reference scale $\mu$ decreases
below $\Lambda$.  

The second possible additional condensation channel is 
\beq 
(N,2) \times (1,2) \to (N,1)
\label{psi_eta_channel_sunxsu2b}
\eeq
with the associated condensate(s)
\beq
\langle\epsilon_{\alpha\beta}\psi^{a,\alpha \ T}_{i,L} C \eta^\beta_{j,L} 
\rangle \ , 
\label{psi_eta_condensate_sunxsu2b}
\eeq
where $1 \le i \le p$ and $1 \le j \le p'$.  Consider the condensates 
(\ref{psi_eta_condensate_sunxsu2b}) with a given $i$, say $i=1$. This set of
condensates (\ref{psi_eta_condensate_sunxsu2b}) breaks SU($N$) to SU($N-p'$) if
$1 \le p' \le N-2$ and and breaks SU($N$) completely if $p' \ge N-1$.  To show
this, note that without loss of generality we may pick $a=N$ and $j=1$ for
one of these condensates.  This condensate, 
$\langle\epsilon_{\alpha\beta}\psi^{N,\alpha \ T}_{1,L} C 
\eta^\beta_{1,L}\rangle$,
breaks SU($N$) to the subgroup SU($N-1$).  The fermions $\psi^{N,\alpha}_{1,L}$
and $\eta^\beta_{1,L}$ involved in this condensate gain dynamical masses of
order the scale at which this condensate forms.  Next, consider the condensate
of the form (\ref{psi_eta_condensate_sunxsu2b}), where now only the SU($N-1$)
gauge indices $a \in \{1,...,N-1\}$ are dynamical.  Again, by convention, we
may pick the SU($N-1$) gauge index in this condensate to be $N-1$ and the copy
index on the $\eta^\beta_{j,L}$ fermion to be $j=2$.  This breaks SU($N-1$) to
SU($N-2$) and the fermions $\psi^{N-1,\alpha}_{1,L}$ and $\eta^\beta_{2,L}$
involved in this condensate gain dynamical masses of order the condensation
scale.  This process continues until SU($N$) is broken to SU($N-p'$) if 
$N-p' \ge 2$ or until SU($N$) is completely broken if $N-p' \le 1$. 
A vacuum alignment argument suggest that it is plausible that 
this pattern of breaking would also hold for other values
$i=2,...,p$.  As noted above, since the SU(2) attractiveness measure of 
all of these condensates, $\Delta C_2=3/2$ is the same, one expects that they
form at essentially the same scale. 

% ==========================================================================

\section{Examination of Some Other ${\rm SU}(N) \otimes {\rm SU}(2)$ 
Theories}
\label{gcaxxu2_section}

Here we examine some $N_G=2$ chiral gauge theories with gauge groups of the
form $G_1 \otimes G_2 = {\rm SU}(N)\otimes {\rm SU}(2)$ in which the $G_1$
sector is of $G_{ca}$ type rather than the $G_{cav}$ type studied in the
previous section.  Two of the simplest cases for the fermion content of the
SU($N$) sector involve chiral fermions transforming according to symmetric and
antisymmetric rank-2 tensor representations of SU($N$), denoted $S_2$ and 
$A_2$, together with the requisite number of fermions in the conjugate
fundamental representation.  Two
minimal anomaly-free SU($N$) sectors are the following,
which we shall label as $S_2 \bar F$ and $A_2 \bar F$:
\beq
S_2 \bar F: \quad \sym + (N+4) \, \overline{\fund} \quad {\rm for} \ N \ge 3
\label{sfbar}
\eeq
and 
\beq
A_2 \bar F: \quad \asym + (N-4) \, \overline{\fund} \quad {\rm for} \ N \ge 5. 
\label{afbar}
\eeq
We restrict the $S_2 \bar F$ theory to have $N \ge 3$, since for $N=2$ it is a
vectorial, rather than chiral, gauge theory.  Similarly, we restrict the 
$A_2 \bar F$ theory to have $N \ge 5$ because for $N=4$, the $\asym$ 
representation is self-conjugate, so the SU(4) $A \bar F$ theory is 
a vectorial, rather than chiral, gauge theory.
Given the contributions to the SU($N$) triangle anomaly from the fermions in
the $S_2$ and $A_2$ representations (see Appendix \ref{group_invariants}),
these respective SU($N$) theories are anomaly-free. 
However, we shall show that neither of these can be used to construct an 
$N_G=2$ direct-product chiral gauge theory in which the SU(2) gauge interaction
is asymptotically free.  

We form the embeddings of the $S_2\bar F$ and 
$A_2\bar F$ sectors in an ${\rm SU}(N) \otimes {\rm SU}(2)$ chiral gauge
theory with the respective fermion contents
\beq
(\sym,2) + (N+4) \, (\overline{\fund},2) \quad {\rm for} \ N \ge 3
\label{sfbarxsu2}
\eeq
and
\beq
(\asym,2) + (N-4) \, (\overline{\fund},2)  \quad {\rm for} \ N \ge 5 \ . 
\label{afbarxsu2}
\eeq
We will denote these as the $S_2 \bar F$ and $A_2 \bar F$
${\rm SU}(N) \otimes {\rm SU}(2)$ theories respectively, and as the 
$T_2 \bar F$ ${\rm SU}(N) \otimes {\rm SU}(2)$ theories (where $T_2$ stands 
for rank-2 tensor) when we refer to them together, with $T_2 = S_2$ or $A_2$. 
These two respective direct-product chiral gauge theories are clearly free of
any anomalies in gauged currents.  With the respective restrictions on $N$,
these theories are of type $(ca,r)$. 

The numbers of SU(2)-doublet fermions in these two respective $T_2 \bar F$ 
${\rm SU}(N) \otimes {\rm SU}(2)$ theories are 
\beq
N_d= \frac{3N(N \pm 3)}{2} \ , 
\label{nd_t2fbarxsu2}
\eeq
where the upper and lower signs refer to the $S_2 \bar F$ and 
$A_2 \bar F$ ${\rm SU}(N) \otimes {\rm SU}(2)$ theories respectively.
In each case, $N_d$ must be even in order for the theory to avoid a global 
SU(2) anomaly.

The one-loop coefficients in the SU($N$) beta function in these respective
theories are 
\beq
b_{1,{\rm SU}(N)} = \frac{1}{3}(7N \mp 12) \ , 
\label{b1_sun_t2fbarxsu2}
\eeq
where again the upper and lower signs refer to the $S_2 \bar F$ and 
$A_2 \bar F$ ${\rm SU}(N) \otimes {\rm SU}(2)$ theories respectively.
In both cases this is positive, so the SU($N$) sector is asymptotically free.

However, the one-loop coefficients in the SU(2) beta function in the respective
theories are 
\beq
b_{1,{\rm SU}(2)} = \frac{1}{3}\Big [ 22 - \frac{3N(N \pm 3)}{2} \Big ] 
\quad {\rm for} \ T_2\bar F \ . 
\label{b1_su2_t2fbarxsu2}
\eeq
We find that for the $S_2\bar F$ ${\rm SU}(N) \otimes {\rm SU}(2)$
theory, $b_{1,{\rm SU}(2)}$ is negative
for all relevant $N \ge 3$.  (with $N$ extended to the positive real numbers,
$b_{1,{\rm SU}(2)} < 0$ for $N > [-9 + \sqrt{609}]/6 = 2.613$), so none of
these theories has the required asymptotically free SU(2) gauge
interaction. Also, many cases are independently excluded by the fact that $N_d$
is odd.  Regarding the $A_2\bar F$ ${\rm SU}(N) \otimes {\rm SU}(2)$ theory,
the $N=5$ case has a positive
$b_{1,{\rm SU}(2)}$ (equal to 7), but is excluded because it has an odd value
of $N_d$, namely $N_d=15$.  All other values of $N$ for the 
$A_2\bar F$ ${\rm SU}(N) \otimes {\rm SU}(2)$ theories are excluded because
$b_{1,{\rm SU}(2)}$ is negative. (With $N$ extended to the positive real
numbers, $b_{1,{\rm SU}(2)} < 0$ for $N > [9+\sqrt{609}]/6=5.613$.)  Many of
these cases are also excluded independently because they have odd $N_d$.
Therefore, our examination of these $T_2\bar F$
${\rm SU}(N) \otimes {\rm SU}(2)$ theories shows that none
of them yields an acceptable chiral gauge theory for our analysis.

% ========================================================================

\section{${\rm SU}(N_c) \otimes {\rm SU}(2)_L \otimes {\rm U}(1)_Y$
Theories}
\label{gsm_section}

Here we shall study the nonperturbative behavior of a chiral gauge theory
with a gauge group of the form (\ref{ggen}) with $N_G=3$, namely 
\beq
G_{GSM}= {\rm SU}(N_c) \otimes {\rm SU}(2)_L \otimes {\rm U}(1)_Y \ , 
\label{gsm}
\eeq
where the subscript GSM stands for ``generalized Standard Model''.  In this
section we will follow an traditional convention in writing some of the fermion
fields as right-handed and, related to this, in denoting the SU(2) gauge group
as SU(2)$_L$.  The fermion content is (with $i=1,...,N_g$, where $N_g=$ number
of generations) 
\beq
Q^{a \alpha}_{i,L} = {u^a_i \choose d^a_i}_L : \quad N_g(N_c,2)_{Y_{Q_L}} 
\label{ql_gsm}
\eeq
(i.e., $Q^{a 1}_{i,L}=u^a_{i,L}$ and $Q^{a 2}_{i,L}=d^a_{i,L}$), 
\beq
q^a_{i,R}, \quad N_g(N_c,1)_{Y_{q_R}} \ , \ q=u,d
\label{qr_gsm}
\eeq
\beq
L^\alpha_{i,L} = {\nu_{\ell_i} \choose \ell_i}_L: \quad N_g(1,2)_{Y_{L_L}} 
\label{ll_gsm}
\eeq
(i.e., $L^1_{i,L}=\nu_{\ell_i,L}$ and $L^2_{i,L}=\ell_{i,L}$), 
\beq
\nu_{\ell_i,R}, \quad N_g(1,1)_{Y_{\nu_R}} \ , 
\label{nur_gsm}
\eeq
and
\beq
\ell_{i,R}: \quad N_g (1,1)_{Y_{\ell_R}} \ .
\label{ellr_gsm}
\eeq
Here, $a$ and $\alpha$ are color and SU(2)$_L$ gauge indices, respectively, and
$i$ is a generational index.  As listed in Table \ref{cgt_types}, this
theory is of type $(cav,r,c)$. For our discussion, we will allow the number of
colors, $N_c$, and $N_g$ and to be arbitrary, subject to the constraints of
asymptotic freedom of the SU($N_c$) and SU(2)$_L$ gauge interactions and the
absence of an SU(2)$_L$ global anomaly. The capital $L$ in Eq. (\ref{ll_gsm})
stands for ``lepton'' and the subscript $L$ for the left-handed chiral
component.  As in the SM, the (chiral) ${\rm SU}(2)_L \otimes {\rm U}(1)_Y$
gauge group contains a vectorial electromagnetic U(1)$_{em}$ subgroup, and the
electric charge satisfies $Q_{em} = T_{3L} + (Y/2)$.  Since
$Q_{em,f_L}=Q_{em,f_R} \equiv Q_{em,f}$ for all fermions $f$, it follows that
the hypercharges of the left-handed and right-handed fermions are related
according to
\beq
Y_{f_R}= 2T_{3L,F_L} + Y_{F_L} \ , 
\label{yqrel_gsm}
\eeq
where here $F$ stands for the left-handed quark or lepton SU(2)$_L$ doublets,
$Q, \ L$. 

This theory is a modification of the Standard Model
with the following changes: (i) the color gauge group is changed from SU(3)$_c$
to ${\rm SU}(N_c)$ and (ii) $N_g$ is arbitrary, both being subject to
the three above-mentioned constraints; (iii) the hypercharge assignments are
generalized from their real-world values, subject to the constraint that there
must not be any gauge anomaly; (iv) two types of 
${\rm SU}(N_c) \otimes {\rm SU}(2)_L$-singlet 
fermions are present, namely $\ell_{i,R}$ and
$\nu_{\ell_i,R}$, are present; and (v) the Higgs scalar boson is removed. The
SU($N_c$) subsector of this theory is vectorial and hence is free of any
anomalies in gauged currents.  As before, the SU(2)$_L$ sector has no pure
cubic SU(2)$_L$ triangle anomaly in gauged currents.  Given the structure of
this GSM theory, the conditions that there be no triangle anomalies in gauged
currents of the form
${\rm SU}(N_c)^2{\rm U}(1)_Y$ and U(1)$_Y^3$ are the same.  If one imposes the
condition that these constraints should be satisfied for each fermion
generation individually, as we will (and as is the case in the SM), then the
resultant condition is
\beq
N_c Y_{Q_L}+Y_{L_L}=0
\label{gsm_anom}
\eeq
for each fermion generation.  The properties of this theory were studied for
the usual case $Y_{L_L}=-1$ in \cite{tmyan} and for general fermion
hypercharge assignments in \cite{nc}.  Provided that the hypercharge
assignments satisfy Eq. (\ref{gsm_anom}), they also yield a vanishing mixed
gravitational-gauge anomaly (for each generation) \cite{nc}. The generic
classes of hypercharge assignments and resultant properties of the theory were
given in \cite{nc}, together with certain special classes.  We comment on these
further below.

The condition that the SU(2)$_L$ gauge sector
should be free of a global anomaly associated with the homotopy group
$\pi_4({\rm SU}(2))={\mathbb Z}_2$ is that the number of SU(2)$_L$ doublets,
\beq
N_d = N_g(N_c+1) 
\label{ndoublets_gsm}
\eeq
is even, i.e., 
\beq
N_g(N_c+1) \ \ {\rm is \ \ even},  
\label{nopi4anom_gsm}
\eeq
and we require that this condition be satisfied. As was noted in \cite{nc}, if
$N_g$ is even, then this constraint allows arbitrary $N_c$, while if
$N_g$ is odd, then it allows only odd $N_c$.  Similarly, if $N_c$ is odd,
then this constrain allows any value of $N_g$, while if $N_c$ is even, it
requires that $N_g$ to be even.

We note that if one were to turn off the U(1)$_Y$ gauge interaction and set
$N_c=N$, then this model would reduce to the special case of Model B in Sect.
\ref{sunxsu2_section} with $p=p'=N_g$ (together with some gauge-singlet
fermions).  The correspondences between fermion fields in these models is
\beq
Q^{a,\alpha}_{i,L} \leftrightarrow \psi^{a,\alpha}_{i,L} \ , \quad 1 \le i \le p
  \ , 
\label{qell_psi_corrspondence}
\eeq
\beq
\{u^a_{i,L}, \ d^a_{i,L} \} \leftrightarrow 
\chi_{a,j,L} \ , \quad 1 \le i \le p;  \ 1 \le j \le 2p
\label{qr_chi_correspondence}
\eeq
and
\beq
L^\alpha_{i,L} \leftrightarrow \eta^\alpha_{i,L} \ , \quad 1 \le i \le p' \ . 
\label{lell_eta_correspondence}
\eeq
One reason that we used abstract notation for the fermions in the Models A, B,
and C of Sect. \ref{sunxsu2_section} is that they have a different structure
than the GSM theory considered here in several respects: (i) the condition
for the absence of anomalies in gauge currents is different, since they have
no U(1) factor; and (ii) $p$ and $p'$ need not be equal,
whereas in the GSM $p=p'=N_g$.  Since the $\nu_{i,R}$ and $\ell_{i,R}$
fields are singlets under ${\rm SU}(N_c) \otimes {\rm SU}(2)_L$, they have no
(nonsinglet) corresponding fields in Model B of Sect. \ref{sunxsu2_section}.

We shall require that both the SU($N_c$) and SU(2)$_L$ gauge interactions in
the GSM must be asymptotically free.  The one-loop coefficient of the SU($N_c$)
beta function is
\beq
b_{1,{\rm SU}(N_c),GSM} = \frac{1}{3}(11N_c-4N_g) \ , 
\label{b1_sunc_gsm}
\eeq
so the requirement that the SU($N_c$) gauge interaction must be 
asymptotically free implies that $N_g$ satisfies 
\beq 
N_g < \frac{11N_c}{4} \ .  
\label{ngen_upper_gsm}
\eeq

The one-loop coefficient of the SU(2)$_L$ beta function is
$b_{1,{\rm SU}(2),GSM} = (1/3)(22-N_d)$, i.e.,
\beq
b_{1,{\rm SU}(2),GSM} = \frac{1}{3}[22-N_g(N_c+1)] \ ,
\label{b1_su2_gsm}
\eeq
so the requirement that the SU(2)$_L$ gauge interaction must be
asymptotically free implies that the number of SU(3)$_L$ doublets 
(\ref{ndoublets_gsm}) is bounded above according to 
\beq
N_g(N_c+1) < 22 \ . 
\label{ngen_ncplus1_upper_gsm}
\eeq
The weak hypercharge U(1)$_Y$ gauge interaction is non-asymptotically free, 
and the associated gauge coupling $g'$ decreases as the Euclidean reference
scale $\mu$ decreases.  If, as we assume, $g'$ is weak at a high scale in the
UV, then it remains weak at lower scales.  Thus, the possible 
nonperturbative behavior in the theory is due to the growth of the gauge
couplings of the non-Abelian gauge interactions.

In our generalized theory, if the SU($N_c$) gauge interaction is sufficiently
strong and dominates over the SU(2)$_L$ interaction, then the former breaks
$G_{EW}$ to U(1)$_{em}$, as in dynamical theories of electroweak symmetry
breaking.  The most attractive channel is
\beq
(\bar N_c,2) \times (N_c,1) \to (1,2) 
\label{mac_gsm_strong_sunc}
\eeq
with attractiveness measure given by (\ref{Delta_C2_ffbar_to1}) with $N=N_c$.
The associated condensates are 
\beq
\langle \bar Q_{a,\alpha,i,L} u^a_{j,R}\rangle
\label{qbar_u_condensate_gsm}
\eeq
and 
\beq
\langle \bar Q_{a,\beta,i,L} d^a_{j,R}\rangle  
\label{qbar_d_condensate_gsm}
\eeq
%\
(and hermitian conjugates). 
With no loss of generality, one may choose $\alpha=1$ in
Eq. (\ref{qbar_u_condensate_gsm}), so that this condensate takes the form
$\langle \bar u_{a,i,L} u^a_{j,R}\rangle$. Since the fermions are massless, one
can order the flavor basis of the $u^a_{j,R}$ fields so that the condensate is
diagonal in flavor and hence has the form  
\beq
\langle \bar u_{a,i,L} u^a_{i,R}\rangle \ , \quad i=1,...,N_g \ . 
\label{ubar_u_condensate_gsm}
\eeq
This condensate thus breaks the electroweak gauge symmetry according to 
${\rm SU}(2)_L \otimes {\rm U}(1)_Y \to {\rm U}(1)_{em}$. 
As noted in \cite{lrs}, a vacuum alignment argument implies that the condensate
(\ref{qbar_d_condensate_gsm}) aligns in a manner so as to preserve this
residual U(1)$_{em}$ gauge symmetry, so that $\beta=2$ in 
(\ref{qbar_d_condensate_gsm}). With an appropriate ordering of the flavor basis
of the $d^a_{j,R}$, this condensate thus takes the form
\beq
\langle \bar d_{a,i,L} d^a_{i,R}\rangle \ , \quad i=1,...,N_g \ . 
\label{dbar_d_condensate_gsm}
\eeq

If, on the other hand, the SU(2)$_L$ gauge interaction is sufficiently strong
and dominates over the SU($N_c$) gauge interaction, then the gauge symmetry
breaking is different.  The most attractive channel for the SU(2)$_L$
interaction is, as before, $2 \times 2 \to 1$.  There are three types of
condensates that can form in this channel, which we denote for short as
$\langle QQ \rangle$, $\langle QL \rangle$, and $\langle LL \rangle$. These
were noted in \cite{smr} for the Standard Model without a Higgs field,
corresponding to the special case of the GSM with $N_c=3$ and $Y_{L_L}=0$.
Here we extend this analysis to the full GSM.  The simplest condensate is
$\langle LL \rangle$, which has the form
\beq
\langle \epsilon_{\alpha\beta}L^{\alpha \ T}_{i,L} C  L^\beta_{j,L}\rangle \ . 
\label{llcondensate_gsm}
\eeq
Using the general property (\ref{symproperty}) and taking into account the
contraction with $\epsilon_{\alpha\beta}$, it follows that the bilinear fermion
operator in (\ref{llcondensate_gsm}) is antisymmetric in the generation indices
$i$ and $j$.  Hence, if $N_g=1$, it is absent.  Assuming $N_g \ge 2$, so that
the condensate (\ref{llcondensate_gsm}) forms, it preserves the 
${\rm SU}(N_c) \otimes {\rm SU}(2)_L$ part of $G_{GSM}$ and, for all but a 
set of measure zero
of hypercharge assignments, it breaks the U(1)$_Y$ weak hypercharge gauge
symmetry, transforming as a $\Delta Y = 2Y_{L_L}$ operator.  The only exception
is the case denoted class ${\rm C2}_{\ell,sym}={\rm C2}_{q,sym}$ in \cite{nc}
(see Tables I and II in \cite{nc}), for which $Y_{L_L}=0=Y_{Q_L}$.  The
condensate (\ref{llcondensate_gsm}) also breaks the (global) lepton family
number U(1)$_{L_i}$ and total lepton number U(1)$_L$ symmetries.  However,
these global symmetries are already broken by SU(2)$_L$ instantons, and
since we assume that SU(2)$_L$ is strongly coupled, these SU(2)$_L$ instantons
are not suppressed as they are (at zero temperature) in the Standard Model.
Note that if one assumes conventional weak hypercharge assignments, so that 
$\nu_{i,R}$ fermions are GSM-singlets, then $\nu_{i,R}^T C \nu_{j,R}$ Majorana
mass terms are, in general, present, and explicitly break
both lepton family number and total lepton number.

The second type of condensate, denoted $\langle QL \rangle$, has the form 
\beq
\langle \epsilon_{\alpha\beta}Q^{a \alpha \ T}_{i,L} C  L^\beta_{j,L}\rangle 
\ . 
\label{qlcondensate_gsm}
\eeq
where $1 \le i,j \le N_g$. This is analogous to the condensate
(\ref{psi_eta_condensate_sunxsu2b}) in the 
${\rm SU}(N) \otimes {\rm SU}(2)_L$ Model B of Section 
\ref{sunxsu2_section}, with the correspondence $N =N_c$ and $p=p'=N_g$, so
our analysis in that section applies here, with these identifications. In
particular, if $N_g \le N_c-2$, then this set of condensates breaks 
${\rm SU}(N_c)$ down to ${\rm SU}(N_c-N_g)$, while if 
$N_g \ge N_c-1$, then this set of condensates breaks 
${\rm SU}(N_c-N_g)$ completely. These condensates also break baryon number and
(total and family) lepton number. 

The third type of condensate, denoted $\langle Q Q \rangle$, has the form 
\beq
\langle \epsilon_{\alpha\beta}Q^{a \alpha \ T}_{i,L} C  Q^{b \beta}_{j,L} 
\rangle \ . 
\label{qqcondensate_gsm}
\eeq
The same analysis that we gave above for the condensate
(\ref{psi_psi_condensate_sunxsu2_strongsu2_sunasym})
in the ${\rm SU}(N) \otimes {\rm SU}(2)$ gauge theory applies here, with 
$N=N_c$ and $p=N_g$  From this analysis we infer that the 
condensation channel is Eq. (\ref{mac_sunxsu2_strongsu2}) with $N=N_c$, 
and the type of $\langle Q Q \rangle$ condensate that is produced here is 
\beqs
& & \langle\epsilon_{\alpha\beta}(
      Q^{\alpha,a \ T}_{i,L} C Q^{\beta,b}_{j,L}
    - Q^{\alpha,b \ T}_{i,L} C Q^{\beta,a}_{j,L} \cr\cr
& +&  Q^{\alpha,a \ T}_{j,L} C Q^{\beta,b}_{i,L}
    - Q^{\alpha,b \ T}_{j,L} C Q^{\beta,a}_{i,L})\rangle \ . 
\label{qqcondensate_gsm_explicit}
\eeqs
This is invariant under SU(2)$_L$ and breaks SU($N_c$) according to
Eq. (\ref{asym_bk_sun}) with $N=N_c$.  For all but a set of measure zero of
weak hypercharge assignments, the condensate (\ref{qqcondensate_gsm_explicit})
also breaks U(1)$_Y$, transforming as a $\Delta Y = 2Y_{Q_L}$ operator.  The
sole exception is the case where $Y_{Q_L}=0=Y_{L_L}$, denoted as class 
${\rm C2}_{q,sym}={\rm C2}_{\ell,sym}$ in \cite{nc} (see Tables I and II in
\cite{nc}).  The condensate (\ref{qqcondensate_gsm_explicit}) also breaks 
baryon number, U(1)$_{B}$, but, as noted, this is already broken by the
SU(2)$_L$ instantons.

% ========================================================================

\section{${\rm SU}(N_c) \otimes {\rm U}(1)_Y$ Theories with $N_c \ge 3$}
\label{sunxu1_section}

In this and the next two sections we shall apply the reduction procedure
discussed in Section \ref{reduction_section} to obtain two (anomaly-free)
$N_G=2$ chiral gauge theories starting with the generalized Standard Model
theory discussed in Section \ref{gsm_section}.  These are obtained by turning
off the SU(2)$_L$ coupling and the SU($N_c$) coupling, respectively.  The third
possibility, namely to turn off the U(1)$_Y$ coupling, yields a theory with the
group ${\rm SU}(N_c) \otimes {\rm SU}(2)_L$, which was already analyzed in
Section \ref{sunxsu2_section}. 

We begin by turning off the SU(2)$_L$ coupling in the generalized Standard
Model, thereby obtaining the gauge group 
\beq
G = {\rm SU}(N_c) \otimes {\rm U}(1)_Y
\label{sunxu1}
\eeq
with the (nonsinglet) fermion content given by Eqs. (\ref{ql_gsm}) and
(\ref{qr_gsm}).  This theory is of the type $(cav,ca)$ in the classification of
Section \ref{methods_section}. As in the GSM itself, because the
SU($N_c$) gauge interaction is vectorial, the ${\rm SU}(N_c)^3$ anomaly is
zero.  In the GSM, $Y_{Q_L}$ denotes the 
generalized weak hypercharge of the left-handed quark doublet in Eq. 
(\ref{ql_gsm}); here, since the theory does not have any SU(2)$_L$, we take it
simply to be the common value of $Y$ for $u^a_{i,L}$ and $d^a_{i,L}$ (and the
same for all $i=1,...,N_g$).  Because the original GSM
contains a vectorial 
${\rm U}(1)_{em} \subset {\rm SU}(2)_L \otimes {\rm U}(1)_Y$, which yields the
relation (\ref{yqrel_gsm}), it follows in the present truncated model that if
we specify $Y_{Q_L}$, then the hypercharges $Y_{u_R}$ and $Y_{d_R}$ are
determined.  Thus, just as was true in the GSM, as discussed in \cite{nc}, in
this truncated version, there is actually an infinite one-parameter family of
models that depend, here, on $Y_Q$.  For any member of this family, as a
special case of the situation in the GSM, it follows that the theory is free of
(i) any ${\rm SU}(N_c)^2{\rm U}(1)_Y$ triangle anomaly, (ii) any U(1)$_Y^3$
anomaly, and (iii) any mixed gravitational-gauge anomaly involving the U(1)$_Y$
gauge group. 

The one-loop coefficient for the ${\rm SU}(N_c)$ beta function is given by
Eq. (\ref{b1_sunc_gsm}), so the upper bound on $N_g$ to ensure the
asymptotic freedom of the SU($N_c$) gauge interaction is the same as in
(\ref{ngen_upper_gsm}).  As the theory evolves from the UV to the IR and the
SU($N_c$) gauge couplings gets sufficiently large, the theory forms bilinear
quark condensates in the SU($N_c$) MAC, which is $\fund \times \overline{\fund}
\to 1$.  {\it A priori}, these condensates would be
\beq
\langle \bar u_{a,i,L} u^a_{j,R}\rangle \ , 
\langle \bar d_{a,i,L} d^a_{j,R}\rangle \ , 
\langle \bar u_{a,i,L} d^a_{j,R}\rangle \ , 
\langle \bar d_{a,i,L} u^a_{j,R}\rangle 
\label{suncxu1}
\eeq
(and their hermitian conjugates).  However, a vacuum alignment argument can be
used to infer that the condensate formation is such as to preserve the
U(1)$_{em}$ subgroup of the U(1)$_Y$ gauge symmetry, i.e., only the $\langle
\bar u_{a,i,L} u^a_{i,R}\rangle$ and $\langle \bar d_{a,i,L} d^a_{i,R}\rangle$
condensates (and their hermitian conjugates) form.  Since the theory has no
bare mass terms, for a fixed ordering of the generational indices of the
left-handed quarks $u^a_{i,L}$ and $d^a_{i,L}$, we can always choose the order
of the the generational indices of the $u^a_{j,R}$ and $d^a_{j,R}$ so that the
condensates are diagonal in generation indices. The $\langle \bar u_{a,i,L}
u^a_{i,R}\rangle$ and $\langle \bar d_{a,i,L} d^a_{i,R}\rangle$ condensates
each break U(1)$_Y$ to U(1)$_{em}$.

% ========================================================================

\section{${\rm SU}(2) \otimes {\rm U}(1)_Y$ Theories}
\label{su2xu1_section}

Here we obtain a chiral gauge theory of the type $(r,ca)$ by starting with 
with the generalized Standard Model of Section \ref{gsm_section} and turning
off the SU($N_c$) gauge coupling, thereby obtaining the gauge group 
\beq
{\rm SU}(2) \otimes {\rm U}(1)_Y \ . 
\label{su2xu1}
\eeq
The fermions are given by (\ref{ql_gsm}) and (\ref{ll_gsm}) of the GSM, with
the modification that now the color index is a global, rather than gauge,
index. The condition that the SU(2) theory must not have any global anomaly is
the same as Eq. (\ref{nopi4anom_gsm}), and, as in the GSM, if one imposes it
individually on each generation, then it is the statement that $N_c$ must be
odd.

The one-loop coefficient in the SU(2)$_L$ beta function is the same as in
Eq. (\ref{b1_su2_gsm}), and the resultant upper bound on $N_g(N_c+1)$ resulting
from the condition that the SU(2)$_L$ gauge interaction must be asymptotically
free is thus the same as in (\ref{ngen_ncplus1_upper_gsm}).  As the theory
evolves from the UV to the IR and the SU(2) grows, if it becomes sufficiently
large, it can produce condensates in the SU(2) MAC, $2 \times 2 \to 1$, of the
three forms discussed in Section \ref{gsm_section}, denoted for short as
$\langle L L \rangle$, $\langle QL \rangle$, and $\langle Q Q \rangle$, with
associated condensates (\ref{llcondensate_gsm}), (\ref{qlcondensate_gsm}), and
(\ref{qqcondensate_gsm_explicit}). As discussed in Section \ref{gsm_section},
except for a set of measure zero, namely the case where $Y_{Q_L}=Y_{L_L}=0$,
denoted ${\rm C2}_{q,sym} = {\rm C2}_{\ell,sym}$ in \cite{nc}, these
condensates break U(1)$_Y$.

% ======================================================================

\section{${\rm SU}(N) \otimes {\rm SU}(2)_L \otimes {\rm SU}(2)_R$ Theories 
with $N \ge 3$}
\label{gn22_section}

In this section we consider another chiral gauge theory with a gauge group of
the form (\ref{ggen}) with $N_G=3$, namely
\beq
G_{N22} = {\rm SU}(N) \otimes {\rm SU}(2)_L \otimes {\rm SU}(2)_R
\label{gn22}
\eeq
with $N \ge 3$ and the fermions
\beq
\psi^{a,\alpha_L}_{i,L}, \ i=1,...,p \ : \quad p \, (\fund,\fund,1) = p \,
(N,2,1) \ , 
\label{psi_ell_gn22}
\eeq
\beq
\psi^{a,\alpha_R}_{i,R}, \ i=1,...,p \ : \quad p \, (\fund,1,\fund) = p \,
(N,1,2) \ , 
\label{psi_r_gn22}
\eeq
\beq
\chi^{\alpha_L}_{j,L}, \ j=1,...,p' \ : \quad p' \, (1,\fund,1) = p' \, (1,2,1)
\ , 
\label{chi_ell_gn22}
\eeq
and
\beq
\chi^{\alpha_R}_{j,R}, \ j=1,...,p'' \ : \quad p'' \, (1,1,\fund) = p'' \,
(1,1,2) \ . 
\label{chi_r_gn22}
\eeq
Here the three representations in the parentheses refer, respectively, to the
three factor groups in Eq. (\ref{gn22}).  As indicated in Table
\ref{cgt_types}, this theory is of type $(cav,r,r)$.  Since the SU($N$) gauge
interaction is vectorial, it has no gauge anomaly, and both the SU(2)$_L$ and
SU(2)$_R$ gauge sectors are safe (anomaly-free).  The conditions that
the SU(2)$_L$ and SU(2)$_R$ gauge sectors should be free of a global anomaly
are, respectively,
\beq
pN+p' \ {\rm is \ even}
\label{nopi4anom_su2ell_gn22}
\eeq
and
\beq
pN+p'' \ {\rm is \ even}. 
\label{nopi4anom_su2r_gn22}
\eeq

As with our other models, we shall require that all three non-Abelian gauge
interactions are asymptotically free.  The one-loop coefficient of the SU($N$)
beta function is the same as in the ${\rm SU}(N) \otimes {\rm SU}(2)_L$ model
of Sect. \ref{sunxsu2_section}, Eq. (\ref{b1_sun_sunxsu2a}) (applicable to both
Models A and B of that section) so the condition that the SU($N$) gauge
interaction should be asymptotically free is the inequality
(\ref{pbound_sunxsu2a_sun_af}).  The one-loop coefficient of the SU(2)$_L$ beta
function is the same as in the ${\rm SU}(N) \otimes {\rm SU}(2)_L$ Model B,
Eq. (\ref{b1_su2_sunxsu2b}), so the requirement that the SU(2)$_L$ gauge
interaction be asymptotically free is the inequality
(\ref{pbound_sunxsu2b_su2_af}).  Finally, the one-loop coefficient of the
SU(2)$_R$ beta function is the same as Eq. (\ref{b1_su2_sunxsu2b}) with $p'$
replaced by $p''$, so the requirement that the SU(2)$_R$ gauge interaction be
asymptotically free is given by the inequality (\ref{pbound_sunxsu2b_su2_af})
with $p'$ replaced by $p''$, namely $Np+p'' < 22$.

We denote the gauge couplings as $g_N$, $g_L$, and $g_R$, with 
$\alpha_N = g_N^2/(4\pi)$, $\alpha_L = g_L^2/(4\pi)$, and  
$\alpha_R = g_R^2/(4\pi)$. If the initial values of these couplings are such
that, as the Euclidean reference scale $\mu$ decreases from large values in
the deep UV, the SU($N$) interaction becomes sufficiently strong and 
dominates over the SU(2)$_L$ and SU(2)$_R$
gauge interactions, then it is expected to produce condensation in 
the most attractive channel, which is 
\beq
(\bar N,2,1) \times (N,1,2) \to (1,2,2) \ , 
\label{mac_gn22}
\eeq
with attractiveness measure (\ref{Delta_C2_ffbar_to1}). 
The associated bilinear fermion condensate is 
\beq
\langle \bar \psi_{a,\alpha_L,i,L} \psi^{a,\alpha_R}_{j,R}\rangle \ . 
\label{psibar_psi_gn22}
\eeq
This breaks ${\rm SU}(2)_L \otimes {\rm SU}(2)_R$ gauge symmetry to the
diagonal (= vector) subgroup, SU(2)$_V$.  That is, if elements of SU(2)$_L$ and
SU(2)$_R$ are denoted as $U_L$ and $U_R$, then SU(2)$_V$ is the subgroup of
${\rm SU}(2)_L \otimes {\rm SU}(2)_R$ defined by the condition $U_L=U_R$.

If the SU(2)$_L$ interaction is sufficiently strong and dominates over both the
SU($N$) and SU(2)$_R$ interaction, then it can produce the three types of
condensates and corresponding symmetry breaking 
discussed in our analysis of the ${\rm SU}(N) \otimes {\rm   SU}(2)_L$ 
Model B above. 

Finally, if the SU(2)$_R$ interaction is sufficiently strong and dominates over
both the SU($N$) and SU(2)$_L$ interaction, our discussion of the condensate
formation in the ${\rm SU}(N) \otimes {\rm   SU}(2)_L$  Model B above applies,
with all subscripts $L$ changed to $R$.

% ========================================================================

\section{${\rm SU}(N_c) \otimes {\rm SU}(2)_L \otimes {\rm SU}(2)_R \otimes {\rm U}(1)_{B-L}$
Theories}
\label{gn221_section}

Here we analyze the chiral gauge theory with a gauge group of 
the form (\ref{ggen}) with $N_G=4$, namely
\beq
G_{N221} = {\rm SU}(N_c) \otimes {\rm SU}(2)_L \otimes {\rm SU}(2)_R \otimes
{\rm U}(1)_{B-L} \ . 
\label{gn221}
\eeq
We denote the gauge couplings as $g_{N_c}$, $g_L$, $g_R$, and $g_U$, with
$\alpha_{N_c}=g_{N_c}^2/(4\pi)$, and so forth for the other couplings.  The 
quarks and leptons in this theory are 
\begin{widetext}
\beq
Q^{a,\alpha_L}_{i,L}, \ i=1,...,N_g \ : 
N_g \, (\fund,\fund,1)_{1/N_c} = N_g \, (N_c,2,1)_{1/N_c} \ , 
\label{q_ell_gn221}
\eeq
\beq
Q^{a,\alpha_R}_{i,R}, \ i=1,...,N_g \ : 
N_g \, (\fund,1,\fund)_{1/N_c} = N_g \, (N_c,1,2)_{1/N_c} \ , 
\label{q_r_gn221}
\eeq
\beq
L^{\alpha_L}_{i,L}, \ i=1,...,N_g \ : 
N_g \, (1,\fund,1)_{-1} = N_g \, (1,2,1)_{-1} \ , 
\label{l_ell_gn221}
\eeq
and
\beq
L^{\alpha_R}_{i,R}, \ i=1,...,N_g \ : 
N_g \, (1,1,\fund)_{-1} = N_g \, (1,1,2)_{-1} \ . 
\label{l_r_gn221}
\eeq
\end{widetext}
Here the three numbers in the parentheses are the dimensionalities of the
SU($N_c$), SU(2)$_L$, and SU(2)$_R$ representations, and the subscripts are the
value of $B-L$, where $B$ and $L$ denote baryon and lepton number.  The capital
$L$ in Eqs. (\ref{l_ell_gn221}) and (\ref{l_r_gn221}) stands for ``lepton'' and
the subscripts $L$ and $R$ for left- and right-handed chiral components, as
before. This theory is of type $(cav,r,r,cav)$ (see Table \ref{cgt_types}) and
is a modification of the model of Ref. \cite{lrs74} in that (i) the number of
colors, $N_c \ge 3$ and (ii) the number of generations, $N_g$, are
arbitrary, subject to constraints to be discussed below; and (iii) the Higgs
field is removed.  One of the interesting features of the original model of
Ref.  \cite{lrs74} is that the $B-L$ operator applied to the full set of quarks
and leptons in each generation has zero trace.  Our generalized model retains
this property, since $B=1/N_c$ for each quark.  A second interesting feature of
the original model is that electric charge $Q_{em}=T_{3L}+T_{3R}+(B-L)/2$ is
quantized, since $T_{3L}$, $T_{3R}$, $B$, and $L$ are rational (indeed, $L$ is
integral).  Again, our generalized model retains this feature.

The SU($N_c$) gauge interaction is vectorial, and hence has no gauge anomaly,
and both the SU(2)$_L$ and SU(2)$_R$ gauge sectors are also free of any pure
cubic gauge anomalies. The theory is also free of 
${\rm SU}(2)_L^2 {\rm U}(1)_{B-L}$, ${\rm SU}(2)_R^2 {\rm U}(1)_{B-L}$, and 
U(1)$_{B-L}^3$ triangle gauge anomalies.  The theory is also free of any mixed
gravitational-gauge anomaly. 
The conditions that the SU(2)$_L$ and SU(2)$_R$ gauge sectors are
each free of any global anomaly are the same, namely the condition
(\ref{nopi4anom_gsm}).

We shall require that the three non-Abelian gauge interactions be
asymptotically free.  The one-loop coefficient of the SU($N_c$) beta function
is the same as in the generalized Standard Model, Eq. (\ref{b1_sunc_gsm}), 
so the condition that the SU($N_c$) gauge interaction must be 
asymptotically free is the same as the inequality (\ref{ngen_upper_gsm}). 
The respective one-loop coefficients of the SU(2)$_L$ and SU(2)$_R$ beta
functions are equal to each other and given by Eq. (\ref{b1_su2_gsm}), 
so the condition that the SU(2)$_L$ and SU(2)$_R$ gauge interactions must be
asymptotically free is the same as the inequality
(\ref{ngen_ncplus1_upper_gsm}).  The U(1)$_{B-L}$ gauge interaction is 
non-asymptotically free, and the associated gauge coupling $g_U$ decreases with
decreasing scale $\mu$.  If, as we assume, $g_U$ is weak at a high scale in the
UV, then it remains weak at lower scales.  Thus, the possible 
nonperturbative behavior in the theory is due to the growth of the gauge
couplings of the three non-Abelian gauge interactions.

If the initial values of these couplings are such
that, as the Euclidean reference scale $\mu$ decreases from large values in
the deep UV, the SU($N_c$) interaction becomes sufficiently strong and 
dominates over the SU(2)$_L$ and SU(2)$_R$
gauge interactions, then it is expected to produce condensation in 
the most attractive channel, which is 
\beq
(\bar N_c,2,1)_{-1/N_c} \times (N_c,1,2)_{1/N_c} \to (1,2,2)_0 \ . 
\label{mac_gn221}
\eeq
The associated bilinear fermion condensate is the same as the one given in Eq.
(\ref{psibar_psi_gn22}). As is evident from (\ref{mac_gn221}), this preserves
the SU($N_c$) and U(1)$_{B-L}$ gauge symmetries and breaks 
${\rm SU}(2)_L \otimes {\rm SU}(2)_R$ to SU(2)$_V$.

If the SU(2)$_L$ interaction is sufficiently strong and dominates over both the
SU($N_c$) and SU(2)$_R$ interaction, then it can produce the three types of
condensates discussed in our analysis of the generalized Standard Model above,
with appropriate changes of weak hypercharge to $B-L$.  The first of these is
the condensate denoted (\ref{llcondensate_gsm}) with the replacements $\alpha,
\ \beta \to \alpha_L, \ \beta_L$ and our discussion in connection with this
condensate applies here. In particular, assuming $N_g \ge 2$, so that this
condensate forms, it preserves the 
${\rm SU}(N_c) \otimes {\rm SU}(2)_L \otimes {\rm SU}(2)_R$ part of 
$G_{N221}$ and breaks the U(1)$_{B-L}$ gauge symmetry,
transforming as $|\Delta L|=2$.

The second type of condensate has the form 
of (\ref{qlcondensate_gsm}) with $i,j=1,...,N_g$. 
This condensate is invariant under the 
${\rm SU}(2)_L \otimes {\rm SU}(2)_R$ part of 
$G_{N221}$ and, for a given $i,j$, it breaks SU($N_c$) to SU($N_c-1$). Without 
loss of generality, we may choose $a=N_c$, so that the residual subgroup 
SU($N_c-1$) operates on the indices $a \in \{1,...,N_c-1\}$.  It also breaks 
U(1)$_{B-L}$, since it transforms as an operator with $|B-L|=|N_c^{-1}-1| 
\ne 0$.

The third type of condensate is $\langle Q Q \rangle$, which has the form 
of (\ref{qqcondensate_gsm}) with $\alpha, \ \beta \to \alpha_L, \ \beta_L$. 
The same analysis that we gave above for this condensate in our discussion of
the generalized Standard Model applies here, with the obvious change of
$\alpha,\beta$ just noted. Thus, again, using MAC and vacuum alignment
arguments, we may infer that the condensate has the explicit structure of 
Eq. (\ref{qqcondensate_gsm_explicit}). This is invariant under the 
${\rm SU}(2)_L \otimes {\rm SU}(2)_R$ part of $G_{N221}$ and breaks 
SU($N_c$) according to Eq. (\ref{asym_bk_sun}) with $N=N_c$.  It also breaks
U(1)$_{B-L}$, transforming as a $|\Delta B|=2/N_c$ operator. 

Finally, if the SU(2)$_R$ interaction is sufficiently strong and dominates over
both the SU($N_c$) and SU(2)$_L$ interaction, then it can produce the three
types of condensates discussed directly above, with the obvious changes of
chiralities of fermion fields from $L$ to $R$ and the resultant changes of
symmetry-breaking patterns.

% ========================================================================

\section{${\rm SU}(N_c+1) \otimes {\rm SU}(2)_L \otimes {\rm SU}(2)_R$ 
Theories}
\label{gpsgen_section}

As noted above, in the original model with an ${\rm SU}(3)_c \otimes {\rm
  SU}(2)_L \otimes {\rm SU}(2)R \otimes {\rm U}(1)_{B-L}$ electroweak gauge
group \cite{lrs74}, the $B-L$ operator applied to the full set of quarks and
leptons in each generation has zero trace. Owing to this property, one can
embed the U(1)$_{B-L}$ gauge symmetry together with SU(3)$_c$ in an SU(4) 
group \cite{ps} such that the
$B-L$ operator ${\rm diag}(1/3,1/3,1/3,-1)$ is proportional to the last
diagonal generator of the Cartan subalgebra of ${\rm su}(4)$.  
The resultant gauge group is 
${\rm SU}(4) \otimes {\rm SU}(2)_L \otimes {\rm SU}(2)_R$.
We may carry out the
same process for our generalized group and thus consider the chiral gauge
theory with gauge group 
\beq
G = {\rm SU}(N_c+1) \otimes {\rm SU}(2)_L \otimes {\rm SU}(2)_R \ . 
\label{ggps}
\eeq
The fermion content is 
\begin{widetext}
\beq
F_L = (Q^{a,\alpha_L}_i, L^{\alpha_L}_i)_L, \ i=1,...,N_g: \quad
N_g \, (\fund,\fund,1) = N_g \, (N_c+1,2,1) \ , 
\label{fl_gps}
\eeq
\beq
F_R = (Q^{a,\alpha_R}_i, L^{\alpha_R}_i)_R, \ i=1,...,N_g: \quad
N_g \, (\fund,1,\fund) = N_g \, (N_c+1,1,2) \ , 
\label{fr_gps}
\eeq
\end{widetext}
where, $a$, $\alpha_L$, and $\alpha_R$ are, respectively, SU($N_c+1$),
SU(2)$_L$, and SU(2)$_R$ gauge indices and $i$ is a generation index. 
The three numbers in the parentheses are the dimensionalities of the
SU($N_c+1$), SU(2)$_L$, and SU(2)$_R$ representations.  The Cartan subalgebra
of ${\rm su}(N_c+1)$ has dimension $N_c+1$ and its last Cartan matrix is
proportional to a diagonal matrix whose first $N_c$ entries are $1/N_c$ and
whose $N_c+1$'th entry is $-1$, i.e., ${\rm diag}(1/N_c,...,1/N_c,-1)$. 

We observe that this model is a special case of the chiral gauge theory that we
analyzed in Section \ref{gn22_section} obtained by setting
$N=N_c+1$, $p=N_g$, $p'=p''=0$, $\psi^{a,\alpha_L}_{i,L} = F_L$, and 
$\psi^{a,\alpha_R}_{i,R} = F_R$.  Thus, this special case of our analysis in
Section \ref{gn22_section} applies for the theory of this section.  

% =======================================================================

\section{${\rm SO}(4k+2) \otimes {\rm SU}(2)$ Theories} 
\label{sonxsu2_section} 

It is also of interest to study chiral gauge theories with direct-product
groups group that involve a safe SO($N$) group. We recall that if $N$ is odd or
if $N$ is even and $N=4k$, $k \ge 1$, then SO($N$) has only real
representations, while if $N=4k+2$ with $k \ge 2$, then the theory has complex
representations but is safe (i.e., has no anomaly for any representation) 
\cite{anomalyfree}.  With this motivation, we consider 
chiral gauge theories with the gauge group
\beq
G = {\rm SO}(4k+2) \otimes {\rm SU}(2) \quad {\rm with} \ k \ge 2 \ . 
\label{gsonxsu2}
\eeq
These are of the form $(cs,cs)$ in the general classification given in Section
\ref{methods_section}. 
Since $N$ is even, it is also convenient to introduce an integer $r=N/2$:
\beq
N=4k+2=2r \ , \quad k \ge 2 \ , 
\label{nrkrel}
\eeq
so $r=2k+1$. As before, we write all fermions as
left-handed.  We start by considering the general fermion content 
\beq
\sum_{{\cal R}, \ {\cal R}'} \Big [ n_{\cal R} \, ({\cal R},1) + 
p \, ({\cal R}',\fund) \Big ] \ , 
\label{sonxsu2_fermions}
\eeq
where ${\cal R}$ and ${\cal R}'$ are representations of SO($4k+2$). 
We include only complex ${\cal R}$ and ${\cal R}'$ since the
use of a real ${\cal R}$ or ${\cal R}'$ would lead to a vectorlike subsector,
so the model would not be irreducibly chiral.  

Using the relevant
group invariants, we calculate the one-loop term in the beta function for the
SO($N$) gauge coupling with $N$ given by (\ref{nrkrel}) to be
\beq
b_{{\rm SO}(4k+2),1} = \frac{2}{3}\Big [ 11(r-1)-\sum_f \Big ( 
n_{\cal R} T_{\cal R} + 2p_{{\cal R}'}T_{{\cal R}'} \Big ) \Big ] \ .
\label{b1_son_sonxsu2}
\eeq
We calculate the one-loop term in the SU(2) beta function to be
\beq
b_{{\rm SU}(2),1} = \frac{1}{3}\Big [ 22 - 2\sum_{{\cal R}'} p_{{\cal R}'} 
{\rm dim}({{\cal R}'}) \Big ] \ . 
\label{b1_su2_sonxsu2}
\eeq
Because the first terms in square brackets in Eq. 
(\ref{b1_son_sonxsu2}) and (\ref{b1_su2_sonxsu2}) are, respectively, linear in
$r$ and a constant, while the relevant $T_{\cal R}$, $T_{{\cal R}'}$, and
${\rm dim}({\cal R}')$ grow exponentially rapidly with $r$, the asymptotic 
freedom of the SO($2r$) and SU(2) gauge interactions places strong 
restrictions on the fermion content and the value of $N$.  For our purposes, it
will be sufficient to consider the simplest models of this type, with 
(complex) ${\cal R}={\cal R}'$. We will consider three specific models, which
we label Models A, B, and C. 

% ==========================================================================

\subsection{ Model A}
\label{sonxsu2a_section}

We first briefly consider the case where the fermion sector has the form 
$\{ f_{ns,s} \}$, i.e, all of the fermions are singlets under SU(2). In this
case, the gauge group effectively reduces to SO($N$), with $N$ given by
(\ref{nrkrel}). We choose the minimal
complex representation for the fermions, namely the 
spinor representation, denoted ${\cal S}$, of dimension
${\rm dim}({\cal S})=2^{r-1}=2^{2k}$ (see Appendix \ref{group_invariants})
and include $n$ copies of these, so the fermion content is 
\beq
\omega_{i,L}, \ \ i=1.,,,n \ : \quad n \, ({\cal S},1) \ , 
\label{omega}
\eeq
where the first and second entries in the parentheses here and below are the
representations of SO($N$) and SU(2), respectively.  The general formula for
the one-loop term in the beta function for the SO($N$) gauge coupling,
Eq. (\ref{b1_son_sonxsu2}) for this Model A reduces to 
\beq
b_{{\rm SO}(2r),1} = \frac{2}{3}\Big [ 11(r-1)-2^{r-4} n \Big ] \ . 
\label{b1_son_sonxsu2a}
\eeq
The requirement that the SO($N$) gauge interaction should be asymptotically
free implies that
\beq
n < \frac{11(r-1)}{2^{r-4}} \ . 
\label{n_upper_sonxsu2a}
\eeq
This has only a finite number of solutions for $n$ that are nontrivial, i.e., 
have $n \ge 1$, and, indeed, also a finite number of solutions for $r$. 
\beq
G_1={\rm SO}(10) \ (i.e., \ k=2, \ r=5) \ \Rightarrow n \le 21
\label{n_upper_so10xsu2a}
\eeq
\beq
G_1={\rm SO}(14) \ (i.e., \ k=3, \ r=7) \ \Rightarrow n \le 8
\label{n_upper_so14xsu2a}
\eeq
\beq
G_1={\rm SO}(18) \ (i.e., \ k=4, \ r=9) \ \Rightarrow n \le 2
\label{n_upper_so18xsu2a}
\eeq
For $k \ge 5$, i.e., $r \ge 11$, the upper bound on $n$ is less than unity, 
precluding any fermions.  

We assume some initial value of the SO($2r$) gauge
coupling in the deep UV and then evolve the theory downward in Euclidean scale
$\mu$.  Recall that the direct product of two spinor representations of 
SO($N$) with $N$ given by (\ref{nrkrel}) is \cite{groupinv} 
\beq
{\cal S} \times {\cal S} = 
2^{2k} \times 2^{2k}=\sum_{\ell=0}^{k-1} A_{2\ell+1} + R_{2k+1;2} 
\ , 
\label{sxsproduct}
\eeq
where $A_t$ denotes the rank-$t$ antisymmetric tensor representation and 
$R_{2k+1}$ is a certain self-dual representation.  The symmetry of the $A_t$
with respect to the interchange of the two spinor representations in the direct
product is given by $(-1)^{u(r,t)}$, where $u(r,t)=(r-t)(r-t-1)/2$
\cite{groupinv}. Thus, for example, one
has, for the lowest relevant value of $k$, namely $k=2$, i.e., 
$G_1={\rm SO}(10)$, 
\beqs
{\rm SO}(10): \quad 
{\cal S} \times {\cal S} & = & 2^4 \times 2^4 = A_1 + A_3 + R_{5;2} \cr\cr
           & = & 10_s + 120_a + 126_s \ , 
\label{sxs_so10}
\eeqs
where the subscripts $s$ and $a$ denote the symmetric and antisymmetric
property of these representations under interchange of the spinors in the
direct product. In general, for SO($2k+2$), from the form of $u(r,t)$, it
follows that $A_1$ is symmetric (resp. antisymmetric) under interchange of the
spinors in the direct product for even $k$ (resp. odd $k$), while $A_3$ is
antisymmetric (resp. symmetric) under interchange of these spinors for even $k$
(resp. odd $k$).

Assuming that the SO($N$) coupling becomes strong enough to produce a
bilinear fermion condensate, the MAC is
\beq
{\rm SO}(N) \ {\rm MAC}: \quad {\cal S} \times {\cal S} \to \fund \ ,
\label{sxs_mac}
\eeq
with attractiveness measure (written, for convenient reference, in terms of
each of the three related parameters $N$, $r$, and $k$) 
\beqs
\Delta C_2 & = & 2C_2({\cal S})-C_2(\fund) = \frac{(N-1)(N-4)}{8} \cr\cr
& = & \frac{(2r-1)(r-2)}{4} = \frac{(4k+1)(2k-1)}{4} \ . \cr\cr
& & 
\label{Deltac2_sxs_to_f_channel}
\eeqs
Since $r \ge 5$, i.e., $k \ge 2$, this is always positive. 
The associated condensate is 
$\langle \omega^T_{i,L} C \omega_{j,L}\rangle$, where
$1 \le i,j \le n$.  From the general result (\ref{symproperty}), it follows 
that the bilinear fermion operator $\omega^T_{i,L} C \omega_{j,L}$ in this
condensate is (i) symmetric under interchange of spinors in the 
${\cal S} \times {\cal S}$ direct product in (\ref{sxs_mac}) 
and hence symmetric in the flavor indices $i,j$ if 
$k$ is even; (ii) antisymmetric under interchange of spinors and hence
antisymmetric in the flavor indices $i,j$ if $k$ is odd. Therefore, explicitly,

\beq
k \ {\rm even} \ \Rightarrow \ 
\langle  \omega^T_{i,L} C \omega_{j,L} + \omega^T_{j,L} C \omega_{i,L}\rangle
\ , \quad 1 \le i,j \le n 
\label{omega_omega_condensate_sonxsu2a_strongson_keven}
\eeq
and
\beq
k \ {\rm odd} \ \Rightarrow \ 
\langle  \omega^T_{i,L} C \omega_{j,L} - \omega^T_{j,L} C \omega_{i,L}\rangle
\ , \quad 1 \le i,j \le n \ . 
\label{omega_omega_condensate_sonxsu2a_strongson_kodd}
\eeq
In both cases, if this condensate forms, then, since it transforms as the
fundamental (vector) representation of the gauge group SO($4k+2$), it breaks 
this symmetry to SO($4k+1$), which is vectorial and does not break further. 

However, if $n=1$ and $k$ is odd, e.g., for SO(14) (i.e., $k=3$), then this
condensate in the MAC channel vanishes identically.  In this case, we consider
the next channel in Eq. (\ref{sxsproduct}), namely
\beq
{\cal S} \times {\cal S} \to A_3
\label{sxs_to_a3}
\eeq
with attractiveness measure 
\beqs
\Delta C_2 & = & 2C_2({\cal S}) - C_2(A_3) = \frac{(N-4)(N-9)}{8} \cr\cr
& = & \frac{(r-2)(2r-9)}{4} = \frac{(2k-1)(4k-7)}{4} \ . 
\cr\cr
& & 
\label{Deltac2_sxs_to_a3_channel}
\eeqs
For the relevant value of $k$, namely $k = 3$, this is $\Delta C_2=25/4$. 

% ==========================================================================

\subsection{ Model B} 
\label{sonxsu2b_section}

Here we consider a model with the gauge group (\ref{gsonxsu2}) with 
(\ref{nrkrel}) and fermion content of the form $\{f_{ns,ns}\}$, namely 
\beq
\psi^\alpha_{i,L}, \ i=1,...,p: \ \  p \, ({\cal S},\fund) \ . 
\label{psi}
\eeq
We denote this as Model B.  Since there are an even number of SU(2) doublets,
this theory has no global SU(2)$_L$ anomaly.

The general formulas for the one-loop coefficients in the SO($N$) 
beta function (with $N$ given by (\ref{nrkrel})) and in the SU(2) beta function
displayed in Eqs. (\ref{b1_son_sonxsu2}) and (\ref{b1_su2_sonxsu2}) reduce, 
for this Model B, to 
\beq
b_{{\rm SO}(2r),1} = \frac{2}{3}[11(r-1)-2^{r-3}p]
\label{b1_son_sonxsu2b}
\eeq
and 
\beq
b_{1,{\rm SU}(2)_L} = \frac{2}{3}(11-2^{r-2}p) \ . 
\label{b1_su2_sonxsu2b}
\eeq
Hence, the respective conditions that the SO($2r$) and SU(2) gauge interactions
should be asymptotically free are
\beq
p < \frac{11(r-1)}{2^{r-3}}
\label{p_upper_sonxsu2b}
\eeq
and
\beq
p < \frac{11}{2^{r-2}} \ . 
\label{pl1_sonxsu2b}
\eeq
Since we take $k \ge 2$, i.e., $r \ge 5$, for our theories, the only possible
nontrivial value for $p$ allowed by the constraint (\ref{pl1_sonxsu2b}) is
$p=1$ and, furthermore, this is only possible for the lowest value of $k$,
namely $k=2$, and thus $G_1={\rm SO}(10)$. No SO($4k+2$) theories of this Model
B type with nonzero fermion content are allowed by the
asymptotic freedom constraint if $k \ge 3$. 

We note that there is consequently no (continuous) nonanomalous global
flavor symmetry of the Lagrangian for this theory.  Since there is only one
copy of the $({\cal S},\fund)$ fermion $\psi^\alpha_{i,L}$, we shall 
henceforth drop
the flavor index and write this field simply as $\psi^\alpha_L$.

If the SO(10) gauge interaction is sufficiently strong and dominates 
over the SU(2) gauge interaction, then it produces a condensate in the 
SO(10) MAC, (\ref{sxs_mac}), thereby breaking the SO(10) gauge 
symmetry to SO(9),
which is vectorial and does not break further. The
condensate is $\langle \psi^{\alpha \ T}_L C \psi^\beta_L \rangle$. As noted
above in Section \ref{sonxsu2a_section}, for SO($4k+2$), the $\fund = A_1$ that
occurs in the Clebsch-Gordan decomposition of the direct product 
${\cal S} \times {\cal S}$ in 
(\ref{sxs_mac}) is symmetric (resp. antisymmetric) under 
interchange of these spinors if $k$ is even (resp. odd).  Since $k=2$ is even
here, it follows that this $\fund$ representation is symmetric under
interchange of the spinors in the direct product.  From the property 
(\ref{symproperty}), it then follows that the SU(2) gauge indices must also be
symmetric, i.e., the SU(2) channel is $2 \times 2 \to 3_s$, so the operator
product transforms as the adjoint (equivalently, the rank-2 symmetric tensor)
representation of SU(2) and hence can be written as proportional to
\beq
\langle \psi^{\alpha \ T}_L C \psi^\beta_L + 
\psi^{\beta \ T}_L C \psi^\alpha_L \rangle \ . 
\label{psi_psi_condensate_sonxsu2b_strongson_keven}
\eeq
Hence, including both factor groups, in this case of a strong and dominant
SO(10) gauge interaction with even $k$ (viz., $k=2$), the condensation is 
in the channel 
\beq
k \ {\rm even} \ \Rightarrow \ 
({\cal S},\fund) \times ({\cal S},\fund) \to (\fund_s,adj_s) = 
((4k+2)_s,3_s) \ . 
\label{psi_psi_channel_sonxsu2b_strongson_keven}
\eeq
In addition to breaking SO(10) to SO(9), this condensate 
SU(2) to a subgroup ${\rm U}(1) \subset {\rm SU}(2)$. 

The $2 \times 2 \to 3_s$ channel is actually a repulsive channel for the SU(2)
interaction, with $\Delta C_2 = -1/2$.  If the SU(2) gauge interaction is weak
enough, this does not matter, but if it is moderately strong, although weaker
than the SO(10) gauge interaction, it might prevent the condensate from
forming.  However, we assume that the SO(10) coupling is sufficiently
strong at a given scale $\mu$ so that this condensate does form. 

Having analyzed the situation in which the SO(10) gauge coupling is strong
and dominates over the SU(2) gauge coupling, we next analyze the opposite
situation in which the SU(2) gauge coupling becomes sufficiently strong and
dominates over the SO(10) coupling. The condensate then forms in the MAC
for SU(2), which is $2 \times 2 \to 1_a$, involving an antisymmetric
contraction of SU(2) indices with the $\epsilon_{\alpha\beta}$ tensor. 
\beq
\langle \epsilon_{\alpha\beta} \psi^{\alpha \ T}_L C \psi^\beta_L \rangle \ . 
\label{psi_psi_condensate_sonxsu2b_strongsu2}
\eeq
The general result (\ref{symproperty}) then implies that the relevant
representation in the Clebsch-Gordan decomposition of the direct product 
${\cal S} \times {\cal S}$ is antisymmetric, and we therefore denote it as 
$R_a$.  
As discussed above, given that $k$ is even here, the representation that would
normally be favored as the MAC in the direct product of two spinors, 
(\ref{sxsproduct}), namely the
$\fund$ representation, is symmetric rather than antisymmetric, and hence 
$R_a$ cannot be $\fund$.  Instead, the
lowest-dimension representation in the expansion (\ref{sxsproduct}) that is odd
under interchange of the spinors is $A_3$ with dimension ${4k+2 \choose 3}$, 
so the condensation channel is 
\beq
({\cal S},\fund) \times ({\cal S},\fund) \to ((A_3)_a,1_a) \ . 
\label{sxs_su2dom_keven}
\eeq
The measure of attractiveness of this channel is given by the $\Delta C_2$ in
Eq. (\ref{Deltac2_sxs_to_a3_channel}) and is always positive for $k \ge 2$.
Explicitly, for our SO(10) Model B theory, the $A_3$ representation has
dimension 120.  When expressed as a sum of product representations of various
SO(10) subgroups, the 120-dimensional representation has no singlets under
either of the maximal (i.e., rank-5) subgroups ${\rm SU}(5) \otimes {\rm U}(1)$
and ${\rm SU}(4) \otimes {\rm SU}(2) \otimes {\rm SU}(2)$, or the rank-4
subgroup SO(9), but does contain a singlet under the rank-4 subgroup 
${\rm SO}(7) \otimes {\rm SU}(2)$ \cite{groupinv}.  
It therefore breaks SO(10) to
${\rm SO}(7) \otimes {\rm SU}(2)$.

% ======================================================================

\subsection{Model C} 
\label{sonxsu2c}

Here we analyze a model, denoted Model C, that has a fermion sector which is a
combination of the fermion sectors of Model A in Section
\ref{sonxsu2a_section} and Model B in Section \ref{sonxsu2b_section}, and
thus is of the form $\{ f_{ns,s}, \ f_{ns,ns} \}$.  These fermions consist of
$n$ copies of the $({\cal S},1)$ fermion $\omega_{i,L}$, $i=1,...,n$, as in
Eq. (\ref{omega}) and a single copy of the $({\cal S},\fund)$ fermion, 
$\psi^\alpha_{1,L}$, as in Eq. (\ref{psi}).

The one-loop coefficient in the beta function of the SU(2) gauge interaction in
this Model C is the same as (\ref{b1_su2_sonxsu2b}) for Model B, and hence 
the requirement that the SU(2) gauge interaction must be asymptotically free
restricts $p \le 1$. The case $p=0$ reduces to Model A, which we have already
discussed.  Therefore, as indicated, we take $p=1$ here.  This, in turn,
restricts $k$ to be equal to 2, i.e., $G_1={\rm SO}(10)$.  

The one-loop coefficient in the SO(10) beta function for this Model C is 
\beq
b_{1,{\rm SO}(10)} = \frac{2}{3}(20-n) \ , 
\label{b1_sonxsu2c}
\eeq
so the asymptotic freedom of the SO(10) gauge interaction implies 
that $n < 20$. 

If the SO(10) gauge interaction is sufficiently strong and dominates over
the SU(2) interaction, then the resultant condensates include those analyzed
for Models A and B above, together with a
new type of condensate.  This new condensate occurs in the channel 
\beq
({\cal S},1) \times ({\cal S},\fund) \to (\fund,\fund)
\label{omega_psi_channel_sonxsu2c}
\eeq
with corresponding condensate
\beq
\langle \omega^T_{i,L} C \psi^\alpha_L\rangle \ , \quad i \in \{1,...,n\} 
\ . 
\label{omega_psi_condensate_sonxsu2c}
\eeq
This condensate breaks SO(10) to SO(9), which is vectorial and does not break
further.

If, on the other hand, the SU(2) gauge interaction is sufficiently strong and
dominates over the SO(10) interaction, then the condensate formation and
symmetry-breaking is the same as for Model B, discussed in Section 
\ref{sonxsu2b_section}. 

% =======================================================================

\section{${\rm SO}(4k+2) \otimes {\rm SU}(M)$ Theory}
\label{sonxsum_section}

Here we consider a chiral gauge theory with the gauge group
\beq
{\rm SO}(N) \otimes {\rm SU}(M) \ , {\rm with} \ N=4k+2=2r \ \ 
 {\rm and} \ M \ge 3 \ . 
\label{sonxsum}
\eeq
We will show that the constraint of asymptotic freedom of both gauge
interactions limits $k$ to the single value $k=2$, but in order to show this,
we must first keep $k \ge 2$ general. The fermion content is the sum over
representations ${\cal R}$ of
\beqs
& & {\rm dim}({\cal R}_{{\rm SU}(M)}) \, ({\cal R}_{{\rm SO}(4k+2)},1) + 
(\bar{\cal R}_{{\rm SO}(4k+2)}, \bar{\cal R}_{{\rm SU}(M)}) \cr\cr
& + & {\rm dim}({\cal R}_{{\rm SO}(4k+2)}) \, (1,{\cal R}_{{\rm SU}(M)}) \ . 
\label{general_fermions_sonxsum}
\eeqs
In the classification of Section {\ref{methods_section}), this theory is of the
  $(cs,cav)$ type.  We take $M \ge 3$ since the theory with $M=2$ has a
  vectorlike subsector comprised of the $(1,2)$ fermions and is
  therefore not irreducibly chiral. Note that even if $M=2$, this theory does
  not coincide with any of Models A, B, or C in Section \ref{sonxsu2_section}
  because those models also avoided $(1,\fund)=(1,2)$ fermions that would have
  constituted a vectorlike subsector.  However, if one were to take $M=2$, then
  the SO($4k+2$)-nonsinglet fermion sector would coincide with that of Model B
  in Section \ref{sonxsu2_section}. We will show below that $M$ is limited to a
  finite set of values by the constraint of asymptotic freedom.  For our
  present purposes, it will suffice to consider the simplest realization of
  this theory, with a single representation ${\cal R}$ of SO($4k+2$), namely
  the smallest complex one, the spinor, and the smallest nonsinglet
  representation of SU(2), namely the fundamental. The resultant fermion
  content is thus
\beq
p \, ({\cal S},\fund) \ , \quad 2^{r-1}p \, (1, \overline{\fund}) \ . 
\label{fermions_sonxsum}
\eeq

The one-loop coefficient of the SO($4k+2$) beta function (with $4k+2=2r$) is
\beq
b_{{\rm SO}(4k+2)} = \frac{2}{3}\Big [ 11(r-1)-2^{r-4}pM \Big ] \ . 
\label{b1_son_sonxsum}
\eeq
The requirement that the SO($4k+2$) gauge interaction must be asymptotically
free then yields the upper bound
\beq
p < \frac{11(r-1)}{2^{r-4}M} \ .
\label{p_upperbound1_sonxsum}
\eeq
Although we restrict $M \ge 3$, we note that if one were to take $M=2$, then
this would be the same as the upper bound (\ref{p_upper_sonxsu2b}) on $p$ for
Model B in Section \ref{sonxsu2_section}). The fact that we take $M \ge 3$ here
makes this a more stringent upper bound than (\ref{p_upper_sonxsu2b}).

We denote the fermion fields for this theory as
\beq
\psi^{\alpha}_{i,L} \ , \quad i=1,...,p: \quad p \, ({\cal S},\fund)
\eeq
and
\beq
\eta_{\alpha,j,L} \ , j = 1,...,2^{r-1}p: \quad 2^{r-1}p \, 
(1,\overline{\fund}), 
\eeq
where $\alpha$ is an SU($M)$ gauge index and $i,j$ are flavor indices. 

The one-loop coefficient of the SU($M$) beta function is
\beq
b_{{\rm SU}(M)} = \frac{1}{3}( 11M - 2^r p) \  . 
\label{b1_sum_sonxsum}
\eeq
The requirement that the SU($M$) gauge interaction must be asymptotically
free then yields the upper bound
\beq
p < \frac{11M}{2^r} \ .
\label{p_upperbound2_sonxsum}
\eeq
For the relevant range $M \ge 3$, these two asymptotic freedom constraints 
can only be satisfied for $r$ equal to its minimal value, $r=5$, i.e., $k=2$
and $G_1={\rm SO}(10)$; furthermore, given that $k=2$, there are only a finite
set of pairs $(M,p)$ that satisfy the constraints.  For the two integer
intervals $3 \le M \le 5$ and $11 \le M \le 21$, only the value $p=1$ is 
allowed, while for $6 \le  M \le 10$, $p$ may take on the values 1 or 2. If 
$M \ge 22$, there are no allowed solutions for $p$.  Our general construction
is thus reduced to the finite family of chiral gauge theories with the gauge
groups ${\rm SO}(10) \otimes {\rm SU}(M)$ with $3 \le M \le 21$ and the
aforementioned possible values of $p$ as a function of $M$.  

If the SO(10) gauge coupling becomes sufficiently large and dominates over the
SU($M$) gauge coupling, then the former can produce condensation in the SO(10)
MAC, namely (\ref{sxs_mac}).  Since the $\fund$ is symmetric under interchange
of the spinors in (\ref{sxs_mac}) for even $k$ and hence, in particular, for
$k=2$, i.e., SO(10), it follows from our general result (\ref{symproperty})
that the combination of the SU($M$) and flavor product $S_{ij}$ must be
symmetric.  For the values of $M$, namely $3 \le M \le 5$ and $11 \le M \le 21$
that allow only $p=1$, it follows that the flavor product must be symmetric, as
$S_{ij}=S_{11}$ and hence that the channel is, in terms of the full
representations, 
\beq
({\cal S},\fund) \times ({\cal S},\fund) \to (\fund_s,\sym)
\label{sxs_sonxsum_strongson}
\eeq
with the condensate 
\beq
\langle \psi^{\alpha \ T}_{1,L} C \psi^{\beta}_{1,L}\rangle 
\label{psi_psi_condensate_sonxsum_strongson_sym}
\eeq
The SO(10) $\Delta C_2$ measure of attractiveness for this channel is given by
the $N=10$ special case of Eq. (\ref{Deltac2_sxs_to_f_channel}), namely 27/4.
However, the SU($M$) $\Delta C_2$ value is negative, as is evident from
Eq. (\ref{Delta_C2_sym_channel}), setting $M=N$, so this is a repulsive channel
as regards the SU($M$) interaction.  This breaks SO(10) to SO(9), which is
vectorial, and does not break further.  Using a vacuum alignment argument, one
may infer that $\alpha=\beta$ so that the condensate
(\ref{psi_psi_condensate_sonxsum_strongson_sym}) breaks SU($M$) to SU($M-1$).

For the interval $6 \le M \le 10$ where the
theory allows $p=2$, the dynamics could instead produce a condensate in the
channel 
\beq
({\cal S},\fund) \times ({\cal S},\fund) \to (\fund_s,\asym)
\label{sxs_sonxsum_strongson_asym}
\eeq
where the flavor product $S_{ij}$ is antisymmetric, so that the condensate is
\beq
 \langle \psi^{\alpha \ T}_{1,L} C \psi^{\beta}_{2,L} 
- \psi^{\alpha \ T}_{2,L} C \psi^{\beta}_{1,L} \rangle \ . 
\label{psi_psi_condensate_sonxsum_strongson_asym}
\eeq
In addition to being attractive as regards the SO(10) interaction, the channel
(\ref{sxs_sonxsum_strongson_asym}) is also attractive with respect to the
SU($M$) interaction, with $\Delta C_2$ given by
Eq. (\ref{Delta_C2_asym_channel}) with $N=M$. Hence, for $M$ in the interval
$6 \le M \le 10$ where $p=2$ is allowed, we infer that the preferred
condensation channel in the case where SO(10) is strong is 
(\ref{sxs_sonxsum_strongson_asym}). This breaks SO(10) to SO(9) and SU($M$) to 
${\rm SU}(M-2) \otimes {\rm SU}(2)$. 

% =====================================================================

\section{${\rm SO}(4k+2) \otimes {\rm SO}(4k'+2)$ Theory}
\label{sonxson_section}

Here we explore a chiral gauge group of the $(cs,cs)$ type, in our
classification from Section (\ref{methods_section}).  For this purpose, we
choose the gauge group
\beq 
{\rm SO}(4k+2) \otimes {\rm SO}(4k'+2) \ , \quad {\rm where} \ k, \ k' \ge 2
\label{gsonson}
\eeq
and fermion content consisting of $p$ copies of the bi-spinor representation, 
$({\cal S},{\cal S})$.  We set $N=4k+2=2r$ and $N'=4k'+2=2r'$. 
Although this family of theories ostensibly depends on the three
parameters $k$, $k'$, and $p$, we will show that there is only one allowed
choice for these three parameters.  

The one-loop coefficients in the SO($4k+2$) and SO($4k'+2$) beta functions are
\beq
b_{{\rm SO}(4k+2),1} = \frac{2}{3}\Big [ 11(r-1)-2^{r+r'-5}p \Big ] 
\label{b1_son_sonxson}
\eeq
and
\beq
b_{{\rm SO}(4k'+2),1} = \frac{2}{3}\Big [ 11(r'-1)-2^{r+r'-5}p \Big ] \ . 
\label{b1_sonp_sonxson}
\eeq
The requirements that the SO($4k+2$) and SO($4k'+2$) gauge interactions must be
asymptotically free yield the upper bounds 
\beq
p < \frac{11(r-1)}{2^{r+r'-5}}
\label{p_upperbound1_sonxson}
\eeq
and
\beq
p < \frac{11(r'-1)}{2^{r+r'-5}}
\label{p_upperbound2_sonxson}
\eeq
These can only be satisfied by the single set of values $r=r'=5$ and $p=1$,
i.e., for the group ${\rm SO}(10) \otimes {\rm SO}(10)$ with $p=1$ copy of the
$({\cal S},{\cal S})$ fermion.  
The structure of this theory is thus symmetric under 
interchange of the two factor groups.  If we break this symmetry by setting one
$\alpha_i$ to be large and the other small in Eq. (\ref{alfsol}), then we can
obtain situations in which one SO(10) coupling dominates over the other.
However, because of the structural symmetry, in contrast to the generic
behavior that we have found for the other direct-product chiral gauge theories
that we have investigated, here the pattern of symmetry breaking is the same
regardless of which SO(10) gauge coupling is large and dominant.  

If the first SO(10) gauge coupling gets large enough and dominates over the
second SO(10) gauge coupling, or vice versa, this can produce fermion 
condensation in the channel 
\beqs
& & ({\cal S},{\cal S}) \times ({\cal S},{\cal S}) \to 
(\fund_s,\fund_s), \ \ i.e., \cr\cr
& & (16,16) \times (16,16) \to (10_s,10_s) 
\label{sxs_sonxson}
\eeqs
where we have used the fact that $k$ and $k'$ are even to infer the symmetry
properties of $(\fund,\fund)$ in the Clebsch-Gordan decomposition of the direct
product of the spinors.  This condensation breaks the gauge symmetry 
${\rm SO}(10) \otimes {\rm SO}(10)$ to 
${\rm SO}(9) \otimes {\rm SO}(9)$, which is vectorial and does not break
further. 

% =====================================================================

\section{${\rm SU}(N) \otimes {\rm SU}(M)$ Theory}
\label{sunxsum_section}

\subsection{General Formulation} 

In this section we analyze a chiral gauge theory with a gauge group
\beq
G = {\rm SU}(N) \otimes {\rm SU}(M)
\label{gsunxsum}
\eeq
and fermion content consisting of a sum over ${\cal R}_{{\rm SU}(N)}$ and 
${\cal R}_{{\rm SU}(M)}$ of 
\beqs
&& {\rm dim}({\cal R}_{{\rm SU}(M)}) \, ({\cal R}_{{\rm SU}(N)},1) + 
( \bar{\cal R}_{{\rm SU}(N)},\bar{\cal R}_{{\rm SU}(M)} ) \cr\cr
& + & {\rm dim}({\cal R}_{{\rm SU}(N)}) \, (1,{\cal R}_{{\rm SU}(M)}) \ ,
\label{fermions_sunxsum}
\eeqs
where ${\cal R}_{{\rm SU}(N)}$ and ${\cal R}_{{\rm SU}(M)}$ denote
representations of SU($N$) and SU($M$), respectively.  This theory is of type
$(cav,cav)$ in the classification of Section \ref{methods_section}. 
A special case of this
theory with ${\cal R}_{{\rm SU}(N)}$ and ${\cal R}_{{\rm SU}(M)}$ both equal to
the fundamental representation was studied before in \cite{georgi86,dfcgt}, 
but in
both of these previous works, it was studied as an example of a preon theory
that might confine without spontaneous symmetry breaking and hence produce
massless composite fermions. Here we consider it in a different way, as a
theory that can self-break with bilinear fermion condensate formation, and we
study the generalized theory with fermion representations higher than the
fundamental. 

The numbers $M \ge 2$ and $N \ge 2$, subject to the asymptotic freedom
constraint (\ref{afgt}) below. This is an irreducibly chiral gauge theory, so
the chiral gauge invariance precludes any mass terms in the fundamental
Lagrangian of the theory. One easily checks that this theory is free of any
anomalies in gauged currents.  It is also free of any global anomalies in the
case where $N$ or $M$ is equal to 2.  To see this, consider, for example, the
case where $N=2$ and the fermions that are nonsinglets under this group
transform as doublets.  From Eq. (\ref{fermions_sunxsum}) one sees that the 
number of SU(2) doublets is $2{\rm dim}({\cal R}_{{\rm SU}(M)})$ and hence is
even. 

We calculate the one-loop coefficients in the SU($N$) and SU($M$) beta
functions to be 
\beqs
& & b_{1,{\rm SU}(N)} = \frac{1}{3}\Big [
11N-4 \, {\dim}({\cal R}_{{\rm SU}(M)}) \, T({\cal R}_{{\rm SU}(N)}) \Big ] 
\cr\cr
& & 
\label{b1_sun_sunxsum}
\eeqs
and
\beqs
& & b_{1,{\rm SU}(M)} = \frac{1}{3}\Big [ 11M-
4 \, {\dim}({\cal R}_{{\rm SU}(N)}) \, T({\cal R}_{{\rm SU}(M)}) \Big ] \ . 
\cr\cr
& & 
\label{b1_sum_sunxsum}
\eeqs
Hence, the requirements that the SU($N$) and SU($M$) gauge interactions should
be asymptotically free imply, respectively, that 
\beq
{\dim}({\cal R}_{{\rm SU}(M)}) \, T({\cal R}_{{\rm SU}(N)}) < \frac{11N}{4} 
\label{dimtrace_upper_sun_af}
\eeq
and
\beq
{\dim}({\cal R}_{{\rm SU}(N)}) \, T({\cal R}_{{\rm SU}(M)}) < \frac{11M}{4} 
\ . 
\label{dimtrace_upper_sum_af}
\eeq
%

% =======================================================================

\subsection{ Model with Fermions $(F,F)$}

Here we consider the version of the general theory of type (\ref{gsunxsum})
containing fermions with ${\cal R}_{{\rm SU}(N)}=\fund$ and 
${\cal R}_{{\rm SU}(M)}=\fund$ (an equivalent notation is $F=\fund$). 
Then
\beq
b_{1,{\rm SU}(N)}=\frac{1}{3}(11N-2M)
\label{b1_sun_sunxsum_ff}
\eeq
and
\beq
b_{1,{\rm SU}(M)}=\frac{1}{3}(11M-2N) \ , 
\label{b1_sum_sunxsum_ff}
\eeq
so the
inequalities (\ref{dimtrace_upper_sun_af}) and (\ref{dimtrace_upper_sum_af})
read $M<11N/2$ and $N<11M/2$, and the range of $N$ and $M$ allowed by these two
constraints is given by
\beq
\frac{2}{11} < \frac{M}{N} < \frac{11}{2} \ . 
\label{afgt}
\eeq
We denote the fermion fields as 
\beq
\omega^a_{i,L} \ , \quad i=1,...,M \ : \quad M \, (N,1) \ , 
\label{omega_gt}
\eeq
\beq
\zeta_{a,\alpha,L} \ : \quad (\bar{N}, \bar M) \ , 
\label{zeta_gt}
\eeq
and 
\beq
\eta^\alpha_{j,L} \ , \quad j=1,...,N \ : \quad N \, (1,M) \ , 
\label{eta_gt}
\eeq
where $a$ and $\alpha$ denote, respectively, SU($N$) and SU($M$) gauge indices
and $i \in \{1,...,M\}$ and and $j \in \{1,...,N\}$ are copy (flavor) indices. 

As noted, one possibility is confinement without any spontaneous chiral
symmetry breaking, leading to massless composite spin 1/2 fermions that are
singlets under ${\rm SU}(N) \otimes {\rm SU}(M)$.  We investigate here the
alternative possibility of condensate formation and associated chiral symmetry
breaking.  If the SU($N$) gauge interaction is sufficiently strong and
dominates over the SU($M$) interaction, then this SU($N$) interaction can
produce condensation in the most attractive channel $N \times \bar N \to 1$.
For the full theory, this is the channel 
\beq
(N,1) \times (\bar N, \bar M) \to (1,\bar M) \ , 
\label{nxnbar_to_1_channel_strongsun}
\eeq
with attractiveness measure given by $\Delta C_2 = 2C_2(N) = (N^2-1)/N$. The 
associated condensates are of the form 
\beq
\langle \omega^{a \ T}_{i,L} C \zeta_{a,\alpha,L}\rangle \ , 
i=1,...,M
\label{omega_zeta_condensate_gt_strongsun}
\eeq
(where the sum over $a$ here and below is from $a=1$ to $a=N$). 
Consider the condensate (\ref{omega_zeta_condensate_gt_strongsun}) with 
$i=1$.  Since this transforms as a $\bar M$ representation of SU($M$), it
breaks this symmetry to SU($M-1$).  By convention, we may use the initial
SU($M$) invariance to pick $\alpha=M$ in
this condensate, so that it is
\beq
\langle \omega^{a \ T}_{1,L} C \zeta_{a,M,L}\rangle \ .
\label{omega_zeta_condensate_gt_strongsun_stage1}
\eeq
We denote the scale where this condensate forms as
$\Lambda$.  The fermions $\omega^a_{1,L}$ and $\zeta_{a,M,L}$ with 
$1 \le a \le N$ involved in this
condensate thus gain dynamical masses of order $\Lambda$, as do the $2M-1$
gauge bosons in the coset ${\rm SU}(M)/{\rm SU}(M-1)$.  In the resultant 
${\rm SU}(N) \otimes {\rm SU}(M-1)$ chiral gauge theory, we consider the
condensate (\ref{omega_zeta_condensate_gt_strongsun}) with $i=2$ and 
$\alpha \in \{1,...,M-1\}$.  Again, by convention, we may use the residual
SU($M-1$) gauge invariance to pick $\alpha=M-1$ in this condensate, so that it
is 
\beq
\langle \omega^{a \ T}_{2,L} C \zeta_{a,M-1,L}\rangle \ .
\label{omega_zeta_condensate_gt_strongsun_stage2}
\eeq
This preserves SU($N$) and transforms like the conjugate fundamental
representation of SU($M-1$), thereby breaking SU($M-1$) to SU($M-2$). 
This fermion condensation process continues with the formation of the 
condensates
\beq
\langle \omega^{a \ T}_{i,L} C \zeta_{a,M-i+1,L}\rangle \ , \quad i \le M \ , 
\label{omega_zeta_condensate_gt_strongsun_stages}
\eeq
breaking SU($M$) completely. The last-enumerated condensate is 
$\langle \omega^{a \ T}_{M,L} C \zeta_{a,1,L}\rangle$. Since all
of the condensates of the form (\ref{omega_zeta_condensate_gt_strongsun}) have
the same attractiveness measure, $\Delta C_2$, they are expected to form at
approximately the same scale, $\Lambda$.  All of the chiral
fermions $\omega^a_{i,L}$ and $\zeta_{a,\alpha}$ with $1 \le i \le M$, 
$1 \le a \le N$, and $1 \le \alpha \le M$ are involved in these condensates and
gain dynamical masses of order $\Lambda$, as do the full set of $M^2-1$ SU($M$)
gauge bosons.  This leaves a theory with an SU($N$) gauge invariance containing
the $N^2-1$ SU($N$) gauge bosons and a set of $MN$ massless SU($N$)-singlet
fermions, namely the $\eta^\alpha_{j,L}$ with $1 \le \alpha \le M$ and 
$1 \le j \le N$.  The SU($N$) pure gluonic theory then forms a spectrum of
SU($N$)-singlet glueballs. 

Clearly, if the SU($M$) gauge interaction is sufficiently strong and dominates
over the SU($N$) gauge interaction, then the above discussion applies with the
replacements $M \leftrightarrow N$ and 
$\omega^a_{i,L} \to \eta^\alpha_{j,L}$. 
In this case, the SU($M$) interaction breaks the SU($N$)
gauge symmetry completely, leaving the $MN$ massless SU($M$)-singlet fermions 
$\omega^a_{i,L}$ with $1 \le a \le N$ and 
$1 \le i \le M$.  The SU($M$) pure gluonic theory then forms a spectrum of
SU($M$)-singlet glueballs. 

The version of the general theory with gauge group (\ref{gsunxsum}) and fermion
representations ${\cal R}_{{\rm SU}(N)}=\fund$ and ${\cal R}_{{\rm
    SU}(M)}=\overline{\fund}$ exhibits the same properties as those that we
have analyzed, with obvious changes, so we do not discuss it separately.

% ========================================================================

\subsection{ Model with $(F,A_2)$}

We next analyze a model with the gauge group (\ref{gsunxsum}) and fermion
representations ${\cal R}_{{\rm SU}(N)}=\fund$ and 
${\cal R}_{{\rm SU}(M)}=\asym$. Since $\asym = \overline{\fund}$ for 
${\rm SU}(M)= {\rm SU}(3)$, we restrict $M \ge 4$.  For this model the 
general equations (\ref{b1_sun_sunxsum}) and (\ref{b1_sum_sunxsum}) read 
\beq
b_{1,{\rm SU}(N)}=\frac{1}{3}[11N-M(M-1)]
\label{b1_sun_sunxsum_fa2}
\eeq
and
\beq
b_{1,{\rm SU}(M)}=\frac{1}{3}[11M-2N(M-2)] \ . 
\label{b1_sum_sunxsum_fa2}
\eeq
The general inequalities (\ref{dimtrace_upper_sun_af}) and
(\ref{dimtrace_upper_sum_af}) guaranteeing the asymptotic freedom of 
the SU($N$) and SU($M$) gauge interactions read, respectively, 
\beq
N > \frac{M(M-1)}{11} 
\label{n_lowerbound_sunxsum_fa2}
\eeq
and
\beq
N < \frac{11M}{2(M-2)}  \ . 
\label{n_upperbound_sunxsum_fa2}
\eeq
\begin{figure}
  \begin{center}
    \includegraphics[height=8cm]{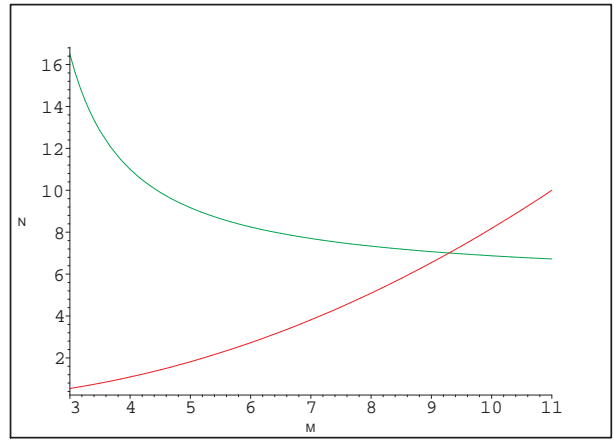}
  \end{center}
\caption{Plot of the region in $M$ and $N$ allowed by the requirement of 
of asymptotic freedom for the SU($N$) and SU($M$) gauge interactions in 
the ${\rm SU}(N) \otimes {\rm SU}(M)$ Model with the $(F,A_2)$ fermion content.
The allowed values of $M$ and $N$ lie between the two curves. 
See text for further discussion.} 
\label{fa2fig}
\end{figure}

In Fig. \ref{fa2fig} we show a plot of the corresponding curves, as a function
of $M$. The lower bound on $N$ from (\ref{n_lowerbound_sunxsum_fa2}) is 
$N > 2$ for $M=4$ and increases as $M$ increases. The upper bound on $N$ from 
(\ref{n_upperbound_sunxsum_fa2}) is $N < 11$ for $M=4$ and decreases as $M$ 
increases.   The curves for the upper and lower bounds on $N$ as a function of
$M$ cross each other at
\beq
M = \frac{3+9\sqrt{3}}{2} = 9.294
\label{msol}
\eeq
where 
\beq
N = \frac{123+18\sqrt{3}}{22} = 7.008 \ , 
\label{nsol}
\eeq
where the floating point values are given to the indicated accuracy. The
allowed values of $M$ and $N$ thus lie within the enclosed region between the
upper and lower curves in Fig. \ref{fa2fig}.  This region has finite area and
hence there are only finitely many allowed values of $M$ and $N$.  This is in
contrast to the joint asymptotic freedom constraint for the model with 
$(F,F)$ fermions, (\ref{afgt}), which is an infinite wedge-shaped region in the
$M$, $N$ plane. As is evident, for a given 
$M \ge 4$, the range of allowed values of $N$ decreases with increasing $M$.
For $M=4$, $N$ may take on values in the range $2 \le N \le 10$, while for
$M=8$, the allowed values of $N$ are $N=6, \ 7$, and for $M=9$, there
is only one allowed value of $N$, namely $N=7$.  If $M \ge 10$, there
are no values of $N$ that satisfy the inequalities 
(\ref{n_lowerbound_sunxsum_fa2}) and (\ref{n_upperbound_sunxsum_fa2}). 

% ==========================================================================

\section{Conclusions}
\label{conclusion_section}

In summary, in this paper we have analyzed patterns of dynamical gauge symmetry
breaking using a variety of chiral gauge theories with direct-product gauge
groups containing asymptotically free non-Abelian gauge interactions of both
unitary and orthogonal types.  Our results on the strong-coupling behavior of
these theories show that these patterns of symmetry breaking are typically
quite different depending on the structure of the factor groups in the direct
product and on which gauge interaction dominates in the formation of fermion
condensates.  These theories provide useful theoretical laboratories
demonstrating explicitly the generic behavior that if the gauge coupling for
one of the factor groups $G_i \subset G$ gets sufficiently strong and dominates
over the other(s), then it can produce bilinear fermion condensates that can
self-break the $G_i$ symmetry itself and/or break other gauge symmetries $G_j
\subset G$. If the $G_i$ gauge interaction that is dominant is vectorial, then
it does not self-break, although it typically still breaks other gauge
symmetries in the direct-product group.  The theories that we have studied also
yield useful examples of sequential gauge symmetry breaking. These results
further elucidate the behavior of strongly coupled chiral gauge theories and
are of value in extending the understanding of nonperturbative behavior of
quantum field theories.

% =======================================================================

\begin{acknowledgments}
This research was partially supported by the NSF grant NSF-PHY-13-16617.
\end{acknowledgments}

% ========================================================================

\begin{appendix}

% Appendix A
\section{Some Relevant Group Invariants}
\label{group_invariants}

For reference, we list some group invariants here. We first define some
notation. Let us denote the generators of the
associated Lie algebra ${\cal g}$ as $T_a$, where $a=1,...,o(G)$, where $o(G)$
is the order of the group.  These generators satisfy the commutation relation
$[T_a,T_b]=ic_{abd}T_d$, where $c_{abc}$ are the structure constants. 
For a representation ${\cal R}$, the Casimir invariants $C_2({\cal R})$ and 
$T({\cal  R})$ are defined as
\beq
\sum_{i,j=1}^{{\rm dim}({\cal R})} 
{\cal D}_{\cal R}(T_a)_{ij} {\cal D}_{\cal R}(T_b)_{ji}=T({\cal R})\delta_{ab} 
\label{tr}
\eeq
and
\beq
\sum_{a=1}^{o(G)} \sum_{j=1}^{{\rm dim}({\cal R})} 
{\cal D}_{\cal R}(T_a)_{ij} {\cal D}_{\cal R}(T_a)_{jk}=
C_2({\cal R})\delta_{ik} \ , 
\label{c2r}
\eeq
where $T_a$ are the generators of $G$, and ${\cal D}_{\cal R}$ is the matrix
representation ({\it Darstellung}) of ${\cal R}$.  These satisfy
\beq
T({\cal R}) \, o(G) = C_2({\cal R}) \, {\rm dim}({\cal R}) \ , 
\label{c2trelation}
\eeq
where ${\rm dim}(R)$ is the dimension of the representation $R$. 

For an SU($N$) group, the rank is $N-1$ and group invariants
(with the normalization convention 
${\rm Tr}(T_aT_b)=(1/2)\delta_{ab}$) include the following
(e.g., \cite{groupinv,rsv99}) 
\beq
C_2(\fund) = \frac{N^2-1}{2N} \ , 
\label{c2fund}
\eeq
\beq
C_2(\sym)=\frac{(N+2)(N-1)}{N} \ , 
\label{c2sym}
\eeq
and
\beq
C_2(\ \asym \ )=\frac{(N-2)(N+1)}{N} \ . 
\label{c2asym}
\eeq

The rank of SO($N$) is the integral part of $N/2$. 
We denote $A_t$ the rank-$t$
antisymmetric tensor representation, with dimension ${N \choose t}$, where
${a \choose b} = a!/[b!(a-b)!]$.  Note that for SO($N$), the adjoint
representation is the same as $A_2$ and the vector, fundamental, and $A_1$ 
representations are the same.  With an appropriate normalization convention
for the generators of SO($N$) (which does not affect the physics), one has
\cite{groupinv,rsv99}
\beq
T(adj) = C_2(adj) = N-2 \ , 
\label{c2g_so}
\eeq
\beq
T({\fund}) = 1 \ , 
\label{t_vector}
\eeq
and
\beq
C_2({\fund}) = \frac{N-1}{2} \ . 
\label{c2_vector}
\eeq
For SO($N$) with $N=2r$ and ${\cal S}$ the spinor representation, 
\beq
{\rm dim}({\cal S})=2^{r-1}
\label{dim_spinor_so_even}
\eeq
\beq
T({\cal S}) = 2^{r-4}
\label{t_spinor_so_neven}
\eeq
\beq
C_2({\cal S}) = \frac{r(2r-1)}{8} \ . 
\label{c2_spinor_so_neven}
\eeq
Denoting the antisymmetric rank-$t$ tensor representation of SO($2r$) as $A_t$,
one has 
\beq
C_2(A_t) = \frac{t(2r-t)}{2} \ . 
\label{c2at}
\eeq

From the structure of the
triangle diagram, it follows that triangle anomaly in gauged currents is
proportional to 
\beq
{\rm Tr}({\cal D}_{\cal R}(T_a),\{ {\cal D}_{\cal R}(T_b), 
{\cal D}_{\cal R}(T_c) \} ) = d_{abc}A_{\cal R}
\label{anomdef}
\eeq
Groups for which $A_{\cal R}=0$ include those with real or pseudoreal
representations, SO($4k+2$) for $k \ge 2$, and E$_6$ 
\cite{groupinv,anomalyfree}.  For the symmetric and
antisymmetric rank-$t$ tensor representations of SU($N$), the anomaly is, 
respectively \cite{anomalyfree} 
\beq
{\cal A}(S_t) = \frac{(N+t)! \, (N+2t)}{(N+2)! \, (t-1)!} \ .
\label{anomskapp}
\eeq
and, for $1 \le t \le N-1$,
\beq
{\cal A}(A_t) = \frac{(N-3)!(N-2t)}{(N-t-1)! (t-1)!}  \ .
\label{anom_asymk}
\eeq
In particular, ${\cal A}(S_2)=N+4$ and ${\cal A}(A_2)=N-4$. 

A gauge theory in $d=4$ dimensions with gauge group $G$ contains instantons if 
$\pi_{d-1}(G)=\pi_3(G)$ is nontrivial.  One has \cite{homotopy}
\beq
\pi_3({\rm SU}(N))={\mathbb Z}
\label{pi3sun}
\eeq
and
\beq
\pi_3({\rm SO}(N)) = {\mathbb Z} \quad {\rm if } \ N \ge 5 \ . 
\label{pi3son}
\eeq
The global anomaly in an SU(2)$_L$ gauge theory is due to 
\beq
\pi_4({\rm SU}(2))={\mathbb Z}_2
\label{pi4su2}
\eeq
Further, 
\beq
\pi_4({\rm SU}(N))= \emptyset \quad {\rm if} \ N \ge 3
\label{pi4sun}
\eeq
and
\beq
\pi_4({\rm SO}(N))= \emptyset \quad {\rm if} \ N \ge 6 \ . 
\label{pi4son}
\eeq
%

% ========================================================================

\end{appendix} 

% ===================================================================

\end{document}